\documentclass[a4paper,twoside,DIV=10]{scrartcl}

\usepackage[T1]{fontenc}
\usepackage[latin1]{inputenc}
\usepackage[english]{babel}
\usepackage{setspace}
\usepackage{titlesec}
\usepackage{amsmath,amssymb,amsthm}
\usepackage{accents}
\usepackage{mathtools}
\usepackage{dsfont}
\usepackage{url}
\usepackage{algpseudocode}
\usepackage{relsize}
\usepackage{array}
\usepackage{dcolumn}
\usepackage{sidecap}
\usepackage{graphicx}
\usepackage[
  algo2e,
  lined,
  boxed,
  commentsnumbered,
  ruled
]{algorithm2e}
\usepackage{xspace}
\usepackage{fancyvrb}
\usepackage{booktabs}
\usepackage[draft]{minted}
\usepackage{subcaption}
\makeatletter
\def\PYG@reset{\let\PYG@it=\relax \let\PYG@bf=\relax%
    \let\PYG@ul=\relax \let\PYG@tc=\relax%
    \let\PYG@bc=\relax \let\PYG@ff=\relax}
\def\PYG@tok#1{\csname PYG@tok@#1\endcsname}
\def\PYG@toks#1+{\ifx\relax#1\empty\else%
    \PYG@tok{#1}\expandafter\PYG@toks\fi}
\def\PYG@do#1{\PYG@bc{\PYG@tc{\PYG@ul{%
    \PYG@it{\PYG@bf{\PYG@ff{#1}}}}}}}
\def\PYG#1#2{\PYG@reset\PYG@toks#1+\relax+\PYG@do{#2}}

\expandafter\def\csname PYG@tok@mf\endcsname{\def\PYG@tc##1{\textcolor[rgb]{0.40,0.40,0.40}{##1}}}
\expandafter\def\csname PYG@tok@gh\endcsname{\let\PYG@bf=\textbf\def\PYG@tc##1{\textcolor[rgb]{0.00,0.00,0.50}{##1}}}
\expandafter\def\csname PYG@tok@cpf\endcsname{\let\PYG@it=\textit\def\PYG@tc##1{\textcolor[rgb]{0.25,0.50,0.50}{##1}}}
\expandafter\def\csname PYG@tok@cp\endcsname{\def\PYG@tc##1{\textcolor[rgb]{0.74,0.48,0.00}{##1}}}
\expandafter\def\csname PYG@tok@sx\endcsname{\def\PYG@tc##1{\textcolor[rgb]{0.00,0.50,0.00}{##1}}}
\expandafter\def\csname PYG@tok@ge\endcsname{\let\PYG@it=\textit}
\expandafter\def\csname PYG@tok@vg\endcsname{\def\PYG@tc##1{\textcolor[rgb]{0.10,0.09,0.49}{##1}}}
\expandafter\def\csname PYG@tok@nc\endcsname{\let\PYG@bf=\textbf\def\PYG@tc##1{\textcolor[rgb]{0.00,0.00,1.00}{##1}}}
\expandafter\def\csname PYG@tok@gd\endcsname{\def\PYG@tc##1{\textcolor[rgb]{0.63,0.00,0.00}{##1}}}
\expandafter\def\csname PYG@tok@vi\endcsname{\def\PYG@tc##1{\textcolor[rgb]{0.10,0.09,0.49}{##1}}}
\expandafter\def\csname PYG@tok@no\endcsname{\def\PYG@tc##1{\textcolor[rgb]{0.53,0.00,0.00}{##1}}}
\expandafter\def\csname PYG@tok@err\endcsname{\def\PYG@bc##1{\setlength{\fboxsep}{0pt}\fcolorbox[rgb]{1.00,0.00,0.00}{1,1,1}{\strut ##1}}}
\expandafter\def\csname PYG@tok@c\endcsname{\let\PYG@it=\textit\def\PYG@tc##1{\textcolor[rgb]{0.25,0.50,0.50}{##1}}}
\expandafter\def\csname PYG@tok@sr\endcsname{\def\PYG@tc##1{\textcolor[rgb]{0.73,0.40,0.53}{##1}}}
\expandafter\def\csname PYG@tok@w\endcsname{\def\PYG@tc##1{\textcolor[rgb]{0.73,0.73,0.73}{##1}}}
\expandafter\def\csname PYG@tok@il\endcsname{\def\PYG@tc##1{\textcolor[rgb]{0.40,0.40,0.40}{##1}}}
\expandafter\def\csname PYG@tok@mb\endcsname{\def\PYG@tc##1{\textcolor[rgb]{0.40,0.40,0.40}{##1}}}
\expandafter\def\csname PYG@tok@kr\endcsname{\let\PYG@bf=\textbf\def\PYG@tc##1{\textcolor[rgb]{0.00,0.50,0.00}{##1}}}
\expandafter\def\csname PYG@tok@kn\endcsname{\let\PYG@bf=\textbf\def\PYG@tc##1{\textcolor[rgb]{0.00,0.50,0.00}{##1}}}
\expandafter\def\csname PYG@tok@ow\endcsname{\let\PYG@bf=\textbf\def\PYG@tc##1{\textcolor[rgb]{0.67,0.13,1.00}{##1}}}
\expandafter\def\csname PYG@tok@go\endcsname{\def\PYG@tc##1{\textcolor[rgb]{0.53,0.53,0.53}{##1}}}
\expandafter\def\csname PYG@tok@mi\endcsname{\def\PYG@tc##1{\textcolor[rgb]{0.40,0.40,0.40}{##1}}}
\expandafter\def\csname PYG@tok@kd\endcsname{\let\PYG@bf=\textbf\def\PYG@tc##1{\textcolor[rgb]{0.00,0.50,0.00}{##1}}}
\expandafter\def\csname PYG@tok@gs\endcsname{\let\PYG@bf=\textbf}
\expandafter\def\csname PYG@tok@ss\endcsname{\def\PYG@tc##1{\textcolor[rgb]{0.10,0.09,0.49}{##1}}}
\expandafter\def\csname PYG@tok@gt\endcsname{\def\PYG@tc##1{\textcolor[rgb]{0.00,0.27,0.87}{##1}}}
\expandafter\def\csname PYG@tok@kt\endcsname{\def\PYG@tc##1{\textcolor[rgb]{0.69,0.00,0.25}{##1}}}
\expandafter\def\csname PYG@tok@sh\endcsname{\def\PYG@tc##1{\textcolor[rgb]{0.73,0.13,0.13}{##1}}}
\expandafter\def\csname PYG@tok@gu\endcsname{\let\PYG@bf=\textbf\def\PYG@tc##1{\textcolor[rgb]{0.50,0.00,0.50}{##1}}}
\expandafter\def\csname PYG@tok@nb\endcsname{\def\PYG@tc##1{\textcolor[rgb]{0.00,0.50,0.00}{##1}}}
\expandafter\def\csname PYG@tok@nf\endcsname{\def\PYG@tc##1{\textcolor[rgb]{0.00,0.00,1.00}{##1}}}
\expandafter\def\csname PYG@tok@m\endcsname{\def\PYG@tc##1{\textcolor[rgb]{0.40,0.40,0.40}{##1}}}
\expandafter\def\csname PYG@tok@se\endcsname{\let\PYG@bf=\textbf\def\PYG@tc##1{\textcolor[rgb]{0.73,0.40,0.13}{##1}}}
\expandafter\def\csname PYG@tok@cm\endcsname{\let\PYG@it=\textit\def\PYG@tc##1{\textcolor[rgb]{0.25,0.50,0.50}{##1}}}
\expandafter\def\csname PYG@tok@c1\endcsname{\let\PYG@it=\textit\def\PYG@tc##1{\textcolor[rgb]{0.25,0.50,0.50}{##1}}}
\expandafter\def\csname PYG@tok@s1\endcsname{\def\PYG@tc##1{\textcolor[rgb]{0.73,0.13,0.13}{##1}}}
\expandafter\def\csname PYG@tok@si\endcsname{\let\PYG@bf=\textbf\def\PYG@tc##1{\textcolor[rgb]{0.73,0.40,0.53}{##1}}}
\expandafter\def\csname PYG@tok@s2\endcsname{\def\PYG@tc##1{\textcolor[rgb]{0.73,0.13,0.13}{##1}}}
\expandafter\def\csname PYG@tok@gr\endcsname{\def\PYG@tc##1{\textcolor[rgb]{1.00,0.00,0.00}{##1}}}
\expandafter\def\csname PYG@tok@sd\endcsname{\let\PYG@it=\textit\def\PYG@tc##1{\textcolor[rgb]{0.73,0.13,0.13}{##1}}}
\expandafter\def\csname PYG@tok@gi\endcsname{\def\PYG@tc##1{\textcolor[rgb]{0.00,0.63,0.00}{##1}}}
\expandafter\def\csname PYG@tok@sc\endcsname{\def\PYG@tc##1{\textcolor[rgb]{0.73,0.13,0.13}{##1}}}
\expandafter\def\csname PYG@tok@sb\endcsname{\def\PYG@tc##1{\textcolor[rgb]{0.73,0.13,0.13}{##1}}}
\expandafter\def\csname PYG@tok@ch\endcsname{\let\PYG@it=\textit\def\PYG@tc##1{\textcolor[rgb]{0.25,0.50,0.50}{##1}}}
\expandafter\def\csname PYG@tok@mo\endcsname{\def\PYG@tc##1{\textcolor[rgb]{0.40,0.40,0.40}{##1}}}
\expandafter\def\csname PYG@tok@ne\endcsname{\let\PYG@bf=\textbf\def\PYG@tc##1{\textcolor[rgb]{0.82,0.25,0.23}{##1}}}
\expandafter\def\csname PYG@tok@s\endcsname{\def\PYG@tc##1{\textcolor[rgb]{0.73,0.13,0.13}{##1}}}
\expandafter\def\csname PYG@tok@kc\endcsname{\let\PYG@bf=\textbf\def\PYG@tc##1{\textcolor[rgb]{0.00,0.50,0.00}{##1}}}
\expandafter\def\csname PYG@tok@nv\endcsname{\def\PYG@tc##1{\textcolor[rgb]{0.10,0.09,0.49}{##1}}}
\expandafter\def\csname PYG@tok@mh\endcsname{\def\PYG@tc##1{\textcolor[rgb]{0.40,0.40,0.40}{##1}}}
\expandafter\def\csname PYG@tok@ni\endcsname{\let\PYG@bf=\textbf\def\PYG@tc##1{\textcolor[rgb]{0.60,0.60,0.60}{##1}}}
\expandafter\def\csname PYG@tok@nl\endcsname{\def\PYG@tc##1{\textcolor[rgb]{0.63,0.63,0.00}{##1}}}
\expandafter\def\csname PYG@tok@k\endcsname{\let\PYG@bf=\textbf\def\PYG@tc##1{\textcolor[rgb]{0.00,0.50,0.00}{##1}}}
\expandafter\def\csname PYG@tok@cs\endcsname{\let\PYG@it=\textit\def\PYG@tc##1{\textcolor[rgb]{0.25,0.50,0.50}{##1}}}
\expandafter\def\csname PYG@tok@na\endcsname{\def\PYG@tc##1{\textcolor[rgb]{0.49,0.56,0.16}{##1}}}
\expandafter\def\csname PYG@tok@o\endcsname{\def\PYG@tc##1{\textcolor[rgb]{0.40,0.40,0.40}{##1}}}
\expandafter\def\csname PYG@tok@bp\endcsname{\def\PYG@tc##1{\textcolor[rgb]{0.00,0.50,0.00}{##1}}}
\expandafter\def\csname PYG@tok@nt\endcsname{\let\PYG@bf=\textbf\def\PYG@tc##1{\textcolor[rgb]{0.00,0.50,0.00}{##1}}}
\expandafter\def\csname PYG@tok@kp\endcsname{\def\PYG@tc##1{\textcolor[rgb]{0.00,0.50,0.00}{##1}}}
\expandafter\def\csname PYG@tok@nn\endcsname{\let\PYG@bf=\textbf\def\PYG@tc##1{\textcolor[rgb]{0.00,0.00,1.00}{##1}}}
\expandafter\def\csname PYG@tok@vc\endcsname{\def\PYG@tc##1{\textcolor[rgb]{0.10,0.09,0.49}{##1}}}
\expandafter\def\csname PYG@tok@gp\endcsname{\let\PYG@bf=\textbf\def\PYG@tc##1{\textcolor[rgb]{0.00,0.00,0.50}{##1}}}
\expandafter\def\csname PYG@tok@nd\endcsname{\def\PYG@tc##1{\textcolor[rgb]{0.67,0.13,1.00}{##1}}}

\def\PYGZbs{\char`\\}
\def\PYGZus{\char`\_}
\def\PYGZob{\char`\{}
\def\PYGZcb{\char`\}}
\def\PYGZca{\char`\^}
\def\PYGZam{\char`\&}
\def\PYGZlt{\char`\<}
\def\PYGZgt{\char`\>}
\def\PYGZsh{\char`\#}
\def\PYGZpc{\char`\%}
\def\PYGZdl{\char`\$}
\def\PYGZhy{\char`\-}
\def\PYGZsq{\char`\'}
\def\PYGZdq{\char`\"}
\def\PYGZti{\char`\~}
\def\PYGZat{@}
\def\PYGZlb{[}
\def\PYGZrb{]}
\makeatother

\newcommand{\periodafter}[1]{#1.}
\titleformat{\section}[runin]{\normalfont\bfseries}{\textbf{\thesection.}}{1ex}{\periodafter}
\titleformat{\subsection}[runin]{\normalfont\bfseries}{\textbf{\thesubsection.}}{1ex}{\periodafter}
\titleformat{\subsubsection}[runin]{\normalfont\bfseries}{\textbf{\thesubsubsection.}}{1ex}{\periodafter}
\titleformat{\paragraph}[runin]{\normalfont\itshape}{}{1ex}{\periodafter}

\date{}

\theoremstyle{plain}

\newcommand{\mycode}[1]{\textnormal{\texttt{#1}}}
\newcommand{\DUNE}{\mycode{DUNE}\xspace}
\newcommand{\dune}[1]{\mycode{dune-#1}}

\newcommand\Cpp{C\nolinebreak[4]\hspace{-.05em}\raisebox{.4ex}{\relsize{-3}{\textbf{++}}}}
\newcommand{\pyMOR}{\mycode{pyMOR}\xspace}
\newcommand{\numpy}{\mycode{NumPy}\xspace}
\newcommand{\python}{\mycode{Python}\xspace}
\newcommand{\Matlab}{\mycode{MATLAB}\xspace}

\renewcommand{\paragraph}[1]{\emph{#1.}}

\newenvironment{code}{\vspace{.5em}\center\BVerbatim}{\endBVerbatim\endcenter\vspace{.5em}}

\title{\Large pyMOR -- Generic Algorithms and Interfaces for Model Order Reduction}

\author{Ren\'e Milk\footnotemark[6] \and Stephan Rave\footnotemark[6] \and Felix Schindler\footnotemark[6]}

\begin{document}
\maketitle

\renewcommand{\thefootnote}{\fnsymbol{footnote}}
\footnotetext[6]{Institute for Computational and Applied Mathematics, University of M\"unster, Einsteinstrasse 62, 48149 M\"unster, Germany, \mycode{\{rene.milk, stephan.rave, felix.schindler\}@uni-muenster.de}}

\begin{small}
\textbf{Abstract.}
Reduced basis methods are projection-based model order reduction techniques for
reducing the computational complexity of solving parametrized partial
differential equation problems. In this work we discuss the design of \pyMOR, a
freely available software library of model order reduction algorithms, in
particular reduced basis methods, implemented with the \python programming
language. As its main design feature, all reduction algorithms in \pyMOR are
implemented generically via operations on well-defined vector array, operator and
discretization interface classes. This allows for an easy integration with existing open-source
high-performance partial differential equation solvers without
adding any model reduction specific code to these solvers. Besides an in-depth
discussion of \pyMOR's design philosophy and architecture, we present several
benchmark results and numerical examples showing the feasibility of our
approach.

\vspace{1ex}


\vspace{1ex}

\end{small}

\pagestyle{myheadings}
\thispagestyle{plain}
\markboth{R.~MILK, S.~RAVE AND F.~SCHINDLER}{GENERIC ALGORITHMS AND
	INTERFACES FOR MODEL ORDER REDUCTION}

\section{Introduction}

Over the past years, model order reduction methods have become an important part
of many numerical simulation workflows for handling large-scale application
problems. Reduced basis (RB) methods are a popular family of such reduction techniques,
applicable to parametrized high-dimensional models described by partial
differential equations (PDEs). The main ingredient of RB methods is a Galerkin projection
of the differential equation onto a problem-adapted reduced subspace generated
from solution snapshots of a high-dimensional approximation of the problem
for certain well-chosen sampling parameters. While the high-dimensional approximation
using standard discretization techniques (such as finite element methods)
often yields discrete function spaces with millions of degrees of freedoms, the
reduced spaces generated by RB methods typically are of order 100 or
smaller, while still retaining the same approximation quality for the problem at hand
as the high-dimensional space. In practice, model order reduction
by RB approximation can lead to speedups of up to several orders of magnitude. By now, a
large body of literature has emerged which theoretically proves and practically
demonstrates the applicability of the RB approach to a large variety of
application problems (see, e.g., the recent monographs
\cite{HesthavenRozzaEtAl2016,QuarteroniManzoniEtAl2016}, the tutorial
\cite{Ha14}, and the references therein).

Despite the popularity of RB methods, only few software
implementations have been discussed in the literature so far. We are only aware of
publications discussing \mycode{rbMIT}~\cite{PateraRozza}, RB modules for
\mycode{libMesh}~\cite{KnezevicPeterson2011} and \mycode{feel++}
\cite{DaversinVeysEtAl2013}, as well as the combination of
\mycode{RBmatlab} with \dune{rb}~\cite{DrohmannHaasdonkEtAl2012a}.
However, these approaches are either too simplistic to be applied to large-scale
problems (\mycode{rbMIT}) or offer limited code re-usability by being tied to a
specific PDE solver ecosystem. While \mycode{RBmatlab}
defines interfaces for the integration with \dune{rb} which would allow to reuse its
algorithms in conjunction with other PDE solvers implementing the same interfaces,
code re-usability is still limited by the fact that \mycode{RBmatlab}
requires all parts of the reduction algorithm involving high-dimensional data to
be implemented by the PDE solver.

In this article we discuss the design of \pyMOR: an open-source, \python-based
model reduction library which is being developed as part of the
\textsc{Multibat} research project (cf.\ \cite{OhlbergerRaveEtAl2014} for a brief overview on
\pyMOR in the context of \textsc{Multibat}). Since one of the goals of said
project is to use RB techniques for the reduction of microscale
battery models implemented independently by another group participating in the
project, an easy integration with third-party PDE solvers is a central design goal
for \pyMOR.

To allow easy integration with external solvers, \pyMOR is built around a small
set of interface classes for interaction with the solver. Unlike the approach taken by
\mycode{RBmatlab}, \pyMOR's interface classes are designed for lower level
communication by directly representing the high-dimensional
vector and operator objects inside the solver. This allows to
implement model order reduction schemes completely within \pyMOR as generic algorithms
operating on these interface classes.
A new external solver is integrated simply by making the solver's data structures
available via a public interface to which \pyMOR can connect to.
No model reduction
specific code has to be added to the solver.
\pyMOR's interface design not only facilitates collaboration between researchers
by decoupling PDE solver and RB algorithm development.
It also fosters the evaluation and adoption of new RB techniques, since algorithms
implemented with \pyMOR can be tested more easily with problems which
have not been considered by the original author.
By now, \pyMOR was used successfully in \cite{BuhrEngwerEtAl2014,OS2015,BEOR15,ORS16}.

A design approach similar to \pyMOR has been taken by the
\mycode{modred}~\cite{BelsonTuEtAl2014} package, a software library implementing
modal decomposition algorithms which operate on generic vector objects
provided by an external source. Due to the algorithmic requirements of
RB methods, however, our approach goes further by also allowing operations on the
solver's system matrices or (nonlinear) operators, resulting in a deeper
integration between the two software components.

Recently, \mycode{RBniCS} was introduced \cite{BallarinRozzaEtAl2015}, a
\python-based RB library built on top of the \mycode{FEniCS} \cite{LoggMardalEtAl2012}
PDE solver library.
Similar to \pyMOR, \mycode{RBniCS} allows easy development of RB applications in
\python while leveraging \mycode{FEniCS} as a high-performance solver.
However, in contrast to \pyMOR, \mycode{RBniCS} is tied to the \mycode{FEniCS}
ecosystem and cannot be integrated with other PDE solvers.
\mycode{redbKIT}, which has been developed as companion software to the recent
monograph \cite{QuarteroniManzoniEtAl2016}, follows a design similar to
\mycode{rbMIT}.
Apart from RB libraries, RB methods are used nowadays in several
specialized simulation softwares such as \mycode{NiftySim} \cite{JohnsenTaylorEtAl2015}
or the commercial code \mycode{Akselos}\footnote{\url{http://www.akselos.com}}.
However, these implementations are restricted to their specific application domain.

This article is organized as follows: Section~\ref{sec:rb_intro} contains a brief
introduction to the RB methodology and provides the mathematical
background needed to follow the technical discussions in the subsequent sections.
We discuss \pyMOR's design from a bird's eye perspective and compare it to
other design approaches in Section~\ref{sec:design}. In Section~\ref{sec:architecture},
we cover \pyMOR's interface classes in more detail, discuss important
implementational aspects and give a basic example, how external solvers
can be integrated with \pyMOR. Moreover, we discuss the parallelization of \pyMOR's reduction
algorithms. In Section~\ref{sec:experiments}, we finally present
technical benchmarks and a more advanced numerical example which demonstrate the performance of
our software design and its applicability to large-scale problems.

\section{The reduced basis method}\label{sec:rb_intro}

In this section we give a very short introduction to the RB
methodology that will hopefully give the reader sufficient background to
understand the discussion of our software design and our numerical experiments.
We mainly focus on the basic class of linear, coercive, affinely decomposed
problems for which the fundamental ideas of the approach can be
most clearly described.
A few of the many extensions of the methodology are discussed in
Section~\ref{sec:rb_extensions}.
For a more detailed introduction to RB methods we refer
to \cite{Ha14,HesthavenRozzaEtAl2016,QuarteroniManzoniEtAl2016}.

\subsection{Linear, coercive, affinely decomposed problems}

We consider pa\-ra\-me\-tri\-zed linear, strongly elliptic problems in weak
form.
More abstractly, we search for solutions $u_\mu\in V$ in some Hilbert space $V$
satisfying
\begin{equation}
	\label{eq:full_problem}
B_\mu(u_\mu, \varphi) = F(\varphi) \qquad \forall \varphi \in V.
\end{equation}
Here, for each parameter $\mu$ contained in some parameter space $\mu \in \mathcal{P}$, $B_\mu$ is a continuous, coercive bilinear form and
$F$ is a continuous linear functional on $V$.
Due to the Lax-Milgram theorem, (\ref{eq:full_problem}) admits a unique solution for every $\mu$.
Moreover, we assume that $B_\mu$ is \emph{affinely decomposed},
i.e., there are $Q \in \mathbb{N}$ bilinear forms $B_1,\ldots,B_Q: V \times V \to \mathbb{R}$,
and mappings $\theta_1,\ldots,\theta_Q: \mathcal{P} \to \mathbb{R}$, such that
\begin{equation}
	\label{eq:affine_decomposition}
	B_\mu = \sum_{q=1}^Q \theta_q(\mu)B_q \quad\forall \mu \in \mathcal{P}.
\end{equation}

Our goal is now to be able to quickly find approximations $u_\mu$ for
arbitrary $\mu \in \mathcal{P}$, assuming we are allowed to compute a few
selected $u_\mu$ during a preceding \emph{offline} phase.
The latter is in practice achieved by assuming that $\ref{eq:full_problem}$ is
already the result of an appropriate high-dimensional discretization of the original
analytical equation, which, however, can only be solved with large computational
effort.

The basic idea of the RB method is to first find
appropriate $\mu_1, \ldots, \mu_N \in \mathcal{P}$ during the offline phase such that $V_N :=
\operatorname{span}\{u_{\mu_1}, \ldots, u_{\mu_N}\}$ is a good approximation space for the so
called solution manifold $\mathcal{M}:= \{u_\mu \ |\ \mu \in \mathcal{P}\}$.
Then, an approximation $u_{N, \mu} \in V_N$ of $u_\mu$ is obtained during the
\emph{online} phase by Galerkin projection
of (\ref{eq:full_problem}), i.e.\ $u_{N, \mu}$ satisfies
\begin{equation}
	\label{eq:reduced_problem}
B_\mu(u_{N,\mu}, \varphi) = F(\varphi) \qquad \forall \varphi \in V_N.
\end{equation}
Again, (\ref{eq:reduced_problem}) has a unique solution according to the
Lax-Milgram theorem.
Assuming, moreover, we have bounds
\begin{equation}
	\alpha_\mu \|\varphi\|^2 \leq B_\mu(\varphi, \varphi) \quad\forall \varphi \in V,
	\qquad\qquad \|B_\mu\| \leq \gamma_\mu \quad\forall \mu \in \mathcal{P},
\end{equation}
Cea's lemma yields the a-priori quasi-best approximation bound
\begin{equation}\label{eq:cea}
	\|u_\mu - u_{N,\mu} \| \leq \frac{\gamma_\mu}{\alpha_\mu} \inf_{\varphi \in V_N}
	\|u_\mu - \varphi\|.
\end{equation}
To solve (\ref{eq:reduced_problem}) numerically, we choose a basis $b_1,\ldots,b_N$ of $V_N$
and let $\underline{u}_{N,\mu} \in \mathbb{R}^N$ be the coefficient vector of
$u_{N, \mu}$ with respect to this basis, i.e.
\begin{equation}\label{eq:reconstruction}
u_{N, \mu} = \sum_{n=1}^N \underline{u}_{N,\mu,n}b_n.
\end{equation}
With $\underline{B}_{N, \mu,i,j} := B_\mu(b_j, b_i)$, $\underline{F}_{N, i}:= F(b_i)$,
$\underline{u}_{N, \mu}$ is then the solution of the linear equation system
$\underline{B}_{N, \mu}\cdot \underline{u}_{N, \mu} = \underline{F}_{N}$.
Due to (\ref{eq:affine_decomposition}), we moreover have $\underline{B}_{N, \mu}
= \sum_{q=1}^Q \theta_q(\mu) \underline{B}_{q, N}$, where $\underline{B}_{q, N}$
denote the matrices of $B_q$. Thus, (\ref{eq:reduced_problem}) can be assembled
and solved with $\mathcal{O}(QN^2 + N^3)$ operations, independently of $\dim V$.
With typical basis sizes of $N \approx 100$, (\ref{eq:reduced_problem}) can then
be solved in less than a millisecond on current hardware.

The coefficients $\underline{u}_{N,\mu}$ usually are
not of interest in themselves. However, if the reduced basis is available, we
can \emph{reconstruct} the reduced solution $u_{N,\mu}$ as an element of the
function space $V$ using (\ref{eq:reconstruction}).
Moreover, one is often only interested in certain quantities of interest which are
derived from the solution $u_\mu$.
If these quantities are linear functionals of $u_\mu$, we can again
pre-compute their evaluations on $b_1, \ldots, b_N$ to arrive at a reduced
model which can produce these quantities completely independent of $\dim V$.

Finally, we have the following standard residual-based a posteriori bound for the
reduction error available %
\begin{equation}
	\label{eq:estimator_efficiency}
	\frac{\alpha_\mu}{\gamma_\mu} \cdot \frac{\| \mathcal{R}_\mu(u_{N,\mu}) \|_{-1}
		}{\alpha_\mu} \leq \|u_\mu - u_{N, \mu}\| \leq \frac{\|
			\mathcal{R}_\mu(u_{N,\mu})\|_{-1}}{\alpha_\mu},
\end{equation}
where the residual $ \mathcal{R}_\mu:V \longrightarrow V^\prime$ is defined by $
\mathcal{R}_\mu[u_{N, \mu}](\varphi) := F(\varphi) - B_\mu(u_{N,\mu}, \varphi)$.
Since we assume $V$ to be finite-dimensional, $\| \mathcal{R}_\mu(u_{N,\mu}) \|_{-1}
$ can be computed as the norm of the Riesz representative
of $\mathcal{R}_\mu(u_{N,\mu})$, given by application of the inverse inner product matrix to
the discrete residual vector.
Using (\ref{eq:affine_decomposition}), this computation can be reduced to
$\mathcal{O}(Q^2N^2)$ online operations.
For cases where a stability estimate $\alpha_\mu$ is not known a priori,
several algorithms have been developed to efficiently compute $\alpha_\mu$
during the online phase (cf.\ \cite[Section 3.7]{QuarteroniManzoniEtAl2016} and references
therein).

\subsection{Basis construction}

As visible from \eqref{eq:cea}, the reduced solution $u_{N,\mu}$ of (\ref{eq:reduced_problem})
is a quasi-optimal approximation of $u_\mu$ within the given reduced
space $V_N$. We are therefore interested in finding spaces $V_N$ minimizing the
maximum best approximation error for $u_\mu$ over all $\mu \in \mathcal{P}$. A
lower bound for what can be achieved is given by the Kolmogorov $N$-width of the
solution manifold $\mathcal{M}$, defined by
\begin{equation}
	d_N( \mathcal{M} ) := \inf_{\substack{V_N \subseteq V \\ \dim V_N \leq N}}
	 \sup_{u \in \mathcal{M}} \inf_{v \in V_N} \|u - v\|.
\end{equation}
For the problem class we consider, it is known that the $N$-widths fall
at least sub-exponentially fast, i.e., there are constants $M,a,\alpha > 0$ s.t.
\begin{equation}
	\label{eq:exponential_decay}
	d_N( \mathcal{M}) \leq Me^{-aN^\alpha}.
\end{equation}
Thus, good approximation spaces $V_N$ do exist.
However, it is usually impossible to find spaces for which this lower bound
$d_N(\mathcal{M})$ is attained.

Reduced basis methods often employ greedy algorithms in order to obtain nearly-best
approximation spaces $V_N$. A standard approach is the error estimator controlled greedy
search defined in Algorithm~\ref{alg:greedy}. This algorithm assumes the
existence of methods \mycode{RBSolve} and \mycode{ErrEst} for solving the reduced
problem and estimating the reduction error given the required reduced model data $ \mathcal{RD}$
which is available through the
\mycode{Reduce} method. In each iteration of the algorithm, the maximum reduction
error over a finite training set $ \mathcal{S}_{train} \subseteq \mathcal{P}$
of parameters is estimated and a high-dimensional solution snapshot $u_{\mu^*}$
for the parameter $\mu^*$ maximizing the error estimates is computed. This
snapshot is then used to extend the reduced space $V_N$.

\begin{algorithm2e}[t]
\DontPrintSemicolon
\KwIn{training set $ \mathcal{S}_{train} \subseteq \mathcal{P}$, tolerance
	$\varepsilon$, max.~dimension $N_{max}$}
\KwOut{reduced spaces $V_1,\ldots,V_N$}
\SetKwFunction{ErrEst}{ErrEst}
\SetKwFunction{RBSolve}{RBSolve}
\SetKwFunction{Solve}{Solve}
\SetKwFunction{Reduce}{Reduce}
$N \leftarrow 0$,\ \ $V_0 \leftarrow \{0\}$,\ \ $ \mathcal{RD} \leftarrow $ \Reduce{$V_N$}\;
\While{$N < N_{max}$ {\bf and} $\max_{\mu \in \mathcal{S}_{train}}$
	\ErrEst{\RBSolve{$\mu$, $ \mathcal{RD}$}, $\mu$, $ \mathcal{RD}$} $> \varepsilon$}{
  $N \leftarrow N + 1$\;
  \nl$\mu^* \leftarrow \operatorname{argmax}_{\mu \in \mathcal{S}_{train}}$	\ErrEst{\RBSolve{$\mu$,
			  $ \mathcal{RD}$}, $\mu$, $ \mathcal{RD}$}\;
  \nl$u_{\mu^*} \leftarrow $ \Solve{$\mu^*$}\;
  \nl$V_{N} \leftarrow \operatorname{span}(V_{N-1} \cup \{u_{\mu^*}\} )$, \ \
	  $\mathcal{RD} \leftarrow$ \Reduce{$V_{N}$}\;
}
\caption{Greedy basis generation with error estimator}
\label{alg:greedy}
\end{algorithm2e}

It is important to note that in Algorithm~\ref{alg:greedy} only lines 2 and 3 of
the main loop involve high-dimensional operations and that these operations are performed only once
per extension step. This allows to afford large training sets $
\mathcal{S}_{train}$, densely sampling the parameter space $ \mathcal{P}$.

If the error estimator used in Algorithm~\ref{alg:greedy} is rigorous and
effective (\ref{eq:estimator_efficiency}), this algorithm is a weak
greedy algorithm in the sense of \cite{BinevCohenEtAl2011}. As a consequence
\cite{DeVorePetrovaEtAl2013}, (\ref{eq:exponential_decay}) carries over to the
resulting spaces $V_N$ in the sense that there are $M^\prime,a^\prime > 0$
only depending on $M$, $a$ and $\sup_{\mu \in \mathcal{P}} \gamma_\mu / \alpha_\mu$, s.t.
\begin{equation}
	\sup_{\mu \in \mathcal{S}_{train}} \inf_{v \in V_N} \|u_\mu - v\| \leq
		M^\prime e^{-a^\prime N^\alpha}.
\end{equation}
In conclusion, if we assume that our training set $\mathcal{S}_{train}$ is
chosen dense enough, we will observe a (sub-)exponentially fast decay of the
maximum reduction error, leading to large computational savings for many application problems.

\subsection{Extensions}\label{sec:rb_extensions}

The presented basic RB methodology can be extended
in many directions, making it suitable
for a wide range of application problems. We briefly cover three extensions
which are relevant for our numerical examples.

\subsubsection{Proper orthogonal decomposition}

As an alternative to a greedy
construction of $V_N$ (Algorithm~\ref{alg:greedy}), we can perform a \lq proper
orthogonal decomposition\rq\ (POD) of a given training set
of solution snapshots $\{u_{\mu_1},\ldots,u_{\mu_K}\} \subset V$ to obtain a set
of orthogonal basis vectors which describe the \lq main directions\rq\ in $V$ by
which the snapshot set is characterized.
More precisely, let $\Phi: \mathbb{R}^K \longrightarrow V$ be the linear
mapping sending the $k$-th canonical basis vector of $\mathbb{R}^K$ to
$u_{\mu_k}$. Then the $k$-th POD mode of the training set is
defined to be the $k$-th left-singular vector of the singular value decomposition of $\Phi$.
The spaces $V_{N, pod}$ spanned by these vectors are $l^2$-optimal in the sense that
\begin{equation}
	\sum_{k=1}^K \inf_{v \in V_{N,pod}} \|u_{\mu,k} - v\|^2 =
	\min_{\substack{V_N \subseteq V \\ \dim V_N \leq N}} \sum_{k=1}^K
	\inf_{v \in V_N} \|u_{\mu,k} - v\|^2.
\end{equation}
POD is therefore a good option if this notion of optimality is desired (and not
the $l^\infty$-optimality the greedy approach is designed for). However, POD
quickly becomes prohibitively expensive when a large training set is needed to
approximate the solution manifold, since the full solution snapshot has to be
computed for each parameter from the training set.

\subsubsection{Instationary problems} The RB methodology can be
extended to instationary problems of the form
\begin{equation}
	\label{eq:full_problem_instationary}
\begin{aligned}
	\langle\partial_t u_\mu(t), \varphi\rangle + B_\mu(u_\mu(t), \varphi) &= F(\varphi) \qquad
\forall \varphi \in V, \\
            u_\mu(0) &= u_{0, \mu}
\end{aligned}
\end{equation}
in a straightforward way. The most common approach is to perform a
Galerkin projection of \eqref{eq:full_problem_instationary} onto a reduced space $V_N \subseteq V$ to arrive at an
ordinary differential equation system of dimension $N$ (method of lines).

Note, however, that solution snapshots are whole trajectories now, so it is no
longer obvious how to extend $V_N$ after a parameter $\mu^*$ has been selected in
Algorithm~\ref{alg:greedy}.
A well-established approach with proven quasi-optimality is the
\textsc{Pod-Greedy} algorithm \cite{Haasdonk2013}, which performs a
POD of the orthogonal projection error trajectory
$u_{\mu^*}(t) - P_{V_N}(u_{\mu^*}(t))$, of which the first modes are then added to
$V_N$.

\subsubsection{Empirical interpolation}\label{sec:ei}
If $B_\mu$ does not satisfy (\ref{eq:full_problem}) or even is nonlinear in
the first variable, the standard RB approach is no longer efficient:
while (\ref{eq:reduced_problem}) is still posed on the low-dimensional space
$V_N$, we cannot solve (\ref{eq:reduced_problem}) online without resorting to high-dimensional
computations which will deteriorate most savings in computation time.

A very generic approach to overcome this issue is empirical operator interpolation: given $x_1,\ldots,x_M \in
V$, $c_1, \ldots, c_M \in V^\prime$ such that the matrix $\underline{I}_M := (c_j(x_i))_{i,j=1}^M$ is
non-singular, we approximate $B_\mu$ by the interpolated form
$\mathcal{I}_M( B_\mu)$ uniquely defined by $\mathcal{I}_M(
B_\mu)(v, \cdot) \in \operatorname{span}\{c_1,\ldots,c_M\}$ and
\begin{equation}
	\mathcal{I}_M(B_\mu)(v, x_i) = B_\mu(v, x_i),\quad i=1,\ldots,M,
\end{equation}
for all $v \in V$. One easily checks that
\begin{equation}
	\label{eq:ei_1}
	\mathcal{I}_M(B_\mu)(v, \cdot) = [c_1,\dots, c_M] \cdot
	\underline{I}_M^{-1} \cdot \big[\,B_\mu(v, x_1), \dots, B_\mu(v, x_M)\,\big]^T.
\end{equation}

In many cases, for instance when the $x_i$ are chosen from a
finite element basis and $B_\mu$ is the finite element
discretization of a partial differential operator, the evaluations
$B_\mu(v, x_1), \ldots, B_\mu(v, x_M)$ can be computed quickly and independently of
$\dim V$ by knowing only $M^\prime < C\cdot M$ degrees of freedom of $v$ for a fixed constant
$C$ (e.g.~depending on the stencil of the discretization).%

Thus, if we approximate $B_\mu$ by $ \mathcal{I}_M( B_\mu)$
in (\ref{eq:reduced_problem}) and substitute (\ref{eq:ei_1}), we obtain
\begin{equation}
	\label{eq:reduced_problem_ei_decomposed}
	[c_1(\varphi),\dots, c_M(\varphi)] \cdot
	\underline{I}_M^{-1} \cdot
	\big[\,B_\mu(u_{N, \mu}, x_1),
	\dots, B_\mu(u_{N, \mu}, x_M)\,\big]^T
	= F(\varphi),
\end{equation}
for all $\varphi \in V_N$,
which can be solved $\dim V$-independently by
pre-computing $c_j(b_n)$ and only storing the coefficients of $b_n$ which are
required for the evaluation of $B_\mu(v, x_m)$.

The \emph{collateral} interpolation basis $c_1, \ldots, c_M$, along with the
interpolation points $x_1$, $\ldots$, $x_M$, is obtained during
the offline phase from a greedy search (\textsc{Ei-Greedy},
\cite{HaasdonkOhlbergerEtAl2008}) or POD (DEIM,
\cite{ChaturantabutSorensen2010}) on an appropriate training set of evaluations
of $B_\mu$.

\section{Design of reduced basis software}\label{sec:design}

In this section we discuss the software design issues which arise when
implementing RB schemes and cover typical design approaches. We then
present the approach taken by \pyMOR and compare it to these standard designs.

\subsection{Required high-dimensional operations}\label{sec:high_dim_operations}

The RB method is by nature a very generic model reduction framework which can
be applied to a wide range of problems. It is an important
feature of RB methods that existing high-dimensional discretizations can be used
as they are in order to derive a reduced order model: any discretization, no
matter if, e.g., continuous finite elements, discontinuous Galerkin or meshless, can be used
as long as the high-dimensional problem is of an appropriate form, such as
(\ref{eq:full_problem}) or (\ref{eq:full_problem_instationary}).

However, while this is true for the mathematical formulation, the code
implementing the high-dimensional discretization nearly always has to be adapted
for model order reduction:

While the main purpose of the solver already is the \emph{computation of
solution snapshots} $u_\mu$, it is often not known a priori (cf.~Algorithm~\ref{alg:greedy})
for which parameters $\mu$ the solution is required. Thus,
either the high-dimensional solver has to stay
initialized in memory during the whole offline phase, or the solver has to be
re-initialized (mesh generation, matrix assembly, etc.) for each new solution snapshot.

To ensure the numerical stability of the reduced model, the new solution
$u_\mu$ has to be orthonormalized w.r.t.\ to the existing reduced basis,
e.g.\ using a stabilized \emph{Gram-Schmidt algorithm}, before it can be added
to the reduced basis.
For instationary problems (\ref{eq:full_problem_instationary}), the solution trajectory has
to be \emph{orthogonally projected} onto the current reduced space $V_N$ and a
\emph{POD} of the projection errors has to be computed.

The \emph{assembly of reduced system matrices} for (\ref{eq:reduced_problem})
requires the evaluation of $B_\mu$ and $F$ for each (combination of) basis
vector(s) of $V_N$. Moreover, the affine decomposition
(\ref{eq:affine_decomposition}) needs to be taken into account for offline/online
decomposition.
The \emph{assembly of the reduction error estimator} requires the
computation of \emph{Riesz representatives} for all residual parts (\ref{eq:affine_decomposition}) and all inner products
between them.
An alternative approach with higher numerical stability \cite{BuhrEngwerEtAl2014}
requires the \emph{computation of an orthonormal basis} for the span of the Riesz representatives
and the
computation of the coefficients of the representatives with respect to the basis.
For the reduction of non-affinely decomposed or nonlinear problems using empirical
operator interpolation (Section~\ref{sec:ei}), $B_\mu$ has to be evaluated on appropriate $u_\mu$,
the collateral basis $c_i$ and interpolation points $x_i$ (\ref{eq:ei_1}) have
to be determined using the \textsc{Ei-Greedy} or \textsc{DEIM} algorithm and
the corresponding reduced data has to be computed.

Finally, if access to the reduced solutions as elements of $V$ is desired (e.g.\
for visualization), the solver needs to be invoked to perform the
\emph{reconstruction} \eqref{eq:reconstruction}.

\subsection{Required low-dimensional operations}\label{sec:low_dim_operations}

In order to \emph{solve the reduced problem} \eqref{eq:reduced_problem},
the same type of algorithms (linear solvers, time-stepping, Newton algorithm)
are required as for the solution of \eqref{eq:full_problem}. However,
the involved data structures will be different: dense instead of sparse matrices,
no or shared memory parallelization instead of distributed memory parallelization.

For reduced problems involving empirical operator interpolation, the \emph{restricted
evaluation} of $B_\mu$ according to (\ref{eq:reduced_problem_ei_decomposed}),
is required.

Lastly, the solution of \eqref{eq:reduced_problem} is already required in the offline
phase during \emph{greedy basis generation} which, in case of adaptive variants
\cite{HaasdonkDihlmannEtAl2011} of Algorithm~\ref{alg:greedy}, may require
substantial implementation work in itself.

\subsection{Classical design approaches}

Reduced basis implementations can be mainly categorized by how they realize
the interaction between high- and low-di\-men\-sio\-nal operations.
All implementations we are aware of fall into one of the following three
categories:

\paragraph{Approach 1: Separate software}
The whole RB scheme is implemented as a single software package which is able to read
and process high-dimensional data produced by some external or
self-written high-dimensional solver.
All high-dimensional operations (Section~\ref{sec:high_dim_operations})
within the offline phase are carried out by the RB software, with the
possible exemption of the solution snapshot computation which may be performed by the solver.
A typical examples is the \mycode{rbMIT}~\cite{PateraRozza} package for \mycode{Matlab}.

The main benefit of this approach is its simplicity: a
single self-contained software package can be developed and maintained by experts
for model order reduction in a programming language ecosystem of their choice.
Interfacing external high-dimensional solvers is simple, only export of system
matrices and (solution) vectors has to be implemented on the solver side.

Consequently, the RB software has to be able to
work with the high-dimensional data produced by the solver. E.g., the given
sparse matrix format has to be supported and an adequate linear solver for the computation of
Riesz representatives needs
to be available.
While this may be easily possible for small to medium sized discretizations
using the tools provided by numerics packages such as \mycode{Matlab}, the
limitations of this approach become obvious when we think of large-scale memory
distributed problems solved on computer clusters where, for instance, a single system
matrix for the whole problem is never assembled. By design, this approach also
cannot be used in conjunction with matrix-free solvers.

Handling of empirical interpolation is problematic, as well: either, for each
model the exact same $B_\mu$ (with correct ordering of degrees of freedom) has to be
re-im\-ple\-men\-ted, or $B_\mu$ has to be evaluated by the external solver. This,
however, does not seem to fit the paradigm of this approach very well.

\paragraph{Approach 2: Inside high-dimensional solver}
The complete reduction process is carried out by the
high-dimensional solver which has been extended by an RB
module. Examples are the RB implementations of
the \mycode{libMesh}~\cite{KnezevicPeterson2011} and \mycode{feel++}
\cite{DaversinVeysEtAl2013} PDE solver packages.

This approach offers the tightest possible integration between the
high-dimensional model and the RB code, allowing maximum
performance and easy development of advanced
reduction techniques which might require special operations on the
high-di\-men\-si\-onal data. Also, there cannot be any
interoperability issues between different versions of the model reduction and
the solver code.

As a downside, implemented algorithms can only be used
within the chosen software ecosystem. Given the large number of available PDE solver
libraries,
this vastly diminishes the reusability of the code and ultimately hinders
collaboration. Moreover, the implementor is required to have a good
understanding of the inner workings of the PDE solver library, which is
typically written in a system language such as \Cpp{}. Many researchers working on
model order reduction do not have such technical knowledge, however.
Consequently many new methods are often only evaluated for ad hoc implementations of academic
\lq toy problems\rq.

\pagebreak

\paragraph{Approach 3: Separate low- and high-dimensional operations}
As a compromise between the aforementioned approaches, all low-dimensional
operations are implemented in a dedicated reduced
basis software which communicates over well-defined interfaces
with the PDE solver carrying out the high-dimensional
operations (Section~\ref{sec:high_dim_operations}). This approach has been
pursued by the integration of \mycode{RBmatlab} with the \dune{rb} module
of the \DUNE numerics environment~\cite{DrohmannHaasdonkEtAl2012a}.

This approach shares many benefits of the
previous two approaches. The main RB logic can be developed
independently from the solver in a programming language of
choice and can be reused with any external solver
implementing the necessary interfaces. All high-dimensional operations are
performed by optimized solver code. Memory distributed or
matrix-free implementations can be easily utilized.

However, extending the PDE solver to perform all
high-dimensional model reduction operations (Section~\ref{sec:high_dim_operations})
still requires substantial work.
A change of the reduction algorithm will usually also require modification of
the RB code in the solver.
This can be a major
issue when both components are developed by separate teams, in particular for research
projects where these components are under
constant change.

\subsection{Design of \pyMOR}

Our design is based on the central observation that all high-dimensional
operations in RB schemes (Section~\ref{sec:high_dim_operations})
can be expressed using only a small set of basic
operations on \emph{operators} and (collections of) \emph{vectors}
which are independent from the concrete reduction scheme in use.
This lead us to the following basic design paradigm for \pyMOR:
\begin{enumerate}
	\item Define model reduction agnostic interfaces for the mathematical
		objects involved with RB methods and related schemes.
	\item Generically implement all algorithms through operations provided
		by these interfaces.
\end{enumerate}

As with design approach 1, \pyMOR offers all model reduction code in a single
self-contained software package.
Since \pyMOR
provides its own basic discretization toolkit, it can be used completely on its
own to easily test new reduction
algorithms. The data types provided by the toolkit can also be used to import
high-dimensional data from disk which has been produced by an external solver.
This allows the use \pyMOR's algorithms in a workflow similar to design
approach 1. Once a new model reduction algorithm has been developed,
\pyMOR's interfaces allow to easily apply the exact same algorithm to
high-fidelity models implemented in a high-performance solver running on a large computer
cluster.

Integrating a new external solver will usually require additional work on the solver side.
However, since \pyMOR's interfaces work on a lower, model reduction agnostic
level compared to design approach 3, these modification can be made mostly independently from the development
of new model reduction algorithms.
Also, our approach offers higher code reusability and
maintainability since the complete reduction algorithm can be implemented in a
single software library. This is particularly important when the model reduction
library shall be used in conjunction with different PDE solver ecosystems.

\pyMOR's interfaces correspond to data structures which are already present in most
PDE solver designs: operators and vectors. Thus,
implementing \pyMOR's interfaces is basically equivalent to exposing the
solver's internal data structures via a public API. Refactoring a PDE solver to offer such an
API generally empowers the
user to easily use and extend the solver in new ways which have not been
envisioned by the developers. E.g., \pyMOR implements
time-stepping algorithms which can be used to easily manufacture
instationary discretizations out of stationary discretizations (see
Section~\ref{sec:burgers}). While time-stepping schemes are
clearly not a focus for \pyMOR's development, a software library of advanced
time-stepping algorithms could use such an API to allow easy testing of these
algorithms with a given discretization, after which a selected algorithm might be
implemented in the solver for maximum performance.
Moreover, a public API also allows
interactive control of the solver, especially when used in conjunction with
dynamic languages such as \python. Such interactive
sessions can be a powerful tool for debugging, allowing to inspect and modify
the solvers state in ways not possible with a classical debugger.

Note that design approach 3 and \pyMOR's design strongly differ in the view on the
relationship between the model reduction software and the high-dimensional solver:
while approach 3 advocates a strong separation between low- and high-dimensional
operations, we think of both components as strongly intertwined. E.g., with
\pyMOR it is easily possible to perform preliminary analyses of reduced models
for which offline/online decomposition is not yet available.

While design approaches 2 and 3 allow to perform all high-dimensional reduction
operations with the maximum efficiency the PDE solver has to offer,
it is clear that \pyMOR's interface design comes at a price of sub-optimal
performance: every call of an interface method incurs a certain overhead,
compiler optimizations may be hindered and the restriction to the available
interface methods may prevent implementations which optimally exploit the given
data structures. To asses the possible loss in performance, we have conducted
several performance benchmarks (Section~\ref{sec:benchmark}) which show that,
while a certain overhead exists, it becomes
negligible for large problems.

\begin{figure}[t]
  \footnotesize%
  \centering%
  \includegraphics[clip, trim=0 586 242 0]{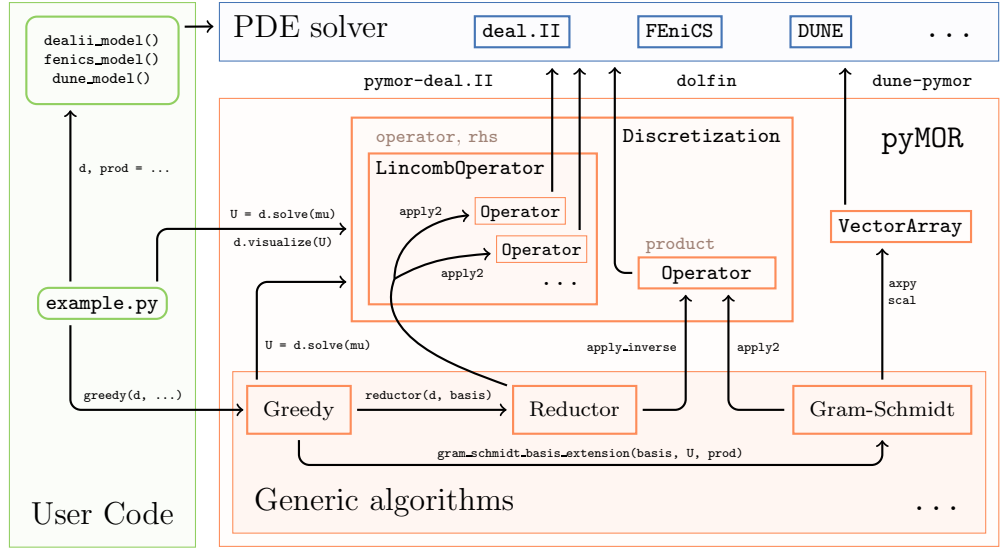}%
  \caption{Schematic view of a typical \pyMOR application displaying the interaction between the user code, external PDE solvers and several parts of \pyMOR (see Section~\ref{sec:example}).}
  \label{fig:interfaces}
\end{figure}

\section{Model Reduction with \pyMOR}\label{sec:architecture}

In this section we present \pyMOR's interfaces in more detail
(Section~\ref{sec:interfaces}) and discuss how the integration of
external solvers via \pyMOR's interfaces can be technically realized
(Section~\ref{sec:python}).
A detailed example for the reduction of models implemented with the
\mycode{FEniCS}, \mycode{deal.II} and \DUNE solver libraries is presented in
Section~\ref{sec:example}.
Finally, we cover the parallelization of \pyMOR's reduction algorithms in
Section~\ref{sec:parallel}.

\subsection{\pyMOR's Interface Classes}\label{sec:interfaces}

\DefineShortVerb{\|}

From a bird's eye perspective, \pyMOR can be seen as a collection of generic
algorithms operating on \mycode{VectorArray}, \mycode{Operator} and
\mycode{Dis}\-\mycode{cre}\-\mycode{ti}\-\mycode{za}\-\mycode{tion} objects which implement the interface methods
that are defined by the abstract \mycode{VectorArrayInterface},
\mycode{OperatorInterface} and \mycode{Discretization} base classes.\footnote{The
full documentation of all interface methods is shipped with \pyMOR and is
available online at \url{http://docs.pymor.org/}.}
To integrate \pyMOR with an external high-dimensional solver, wrapper classes for these types
representing the corresponding high-dimensional data
inside the solver have to be implemented
(cf.\ Figure~\ref{fig:interfaces}).

\begin{listing}[t]
\caption{Main script of \texttt{example.py} (simplified).}
\label{alg:example_main}
\begin{Verbatim}[commandchars=\\\{\}]
\PYG{k+kn}{from} \PYG{n+nn}{pymor.basic} \PYG{k+kn}{import} \PYG{o}{*}
\PYG{k+kn}{from} \PYG{n+nn}{functools} \PYG{k+kn}{import} \PYG{n}{partial}

\PYG{n}{d}\PYG{p}{,} \PYG{n}{prod} \PYG{o}{=} \PYG{n}{pymor\PYGZus{}model}\PYG{p}{()}

\PYG{n}{coerc\PYGZus{}est} \PYG{o}{=} \PYG{n}{ExpressionParameterFunctional}\PYG{p}{(}\PYG{l+s+s1}{\PYGZsq{}min(diffusion)\PYGZsq{}}\PYG{p}{,} \PYG{n}{d}\PYG{o}{.}\PYG{n}{parameter\PYGZus{}type}\PYG{p}{)}
\PYG{n}{reductor} \PYG{o}{=} \PYG{n}{partial}\PYG{p}{(}\PYG{n}{reduce\PYGZus{}stationary\PYGZus{}coercive}\PYG{p}{,}
                   \PYG{n}{error\PYGZus{}product}\PYG{o}{=}\PYG{n}{prod}\PYG{p}{,} \PYG{n}{coercivity\PYGZus{}estimator}\PYG{o}{=}\PYG{n}{coerc\PYGZus{}est}\PYG{p}{)}
\PYG{n}{rd}\PYG{p}{,} \PYG{n}{rc} \PYG{o}{=} \PYG{n}{reduce\PYGZus{}naive}\PYG{p}{(}\PYG{n}{d}\PYG{p}{,} \PYG{n}{reductor}\PYG{p}{,} \PYG{l+m+mi}{10}\PYG{p}{)}

\PYG{n}{res} \PYG{o}{=} \PYG{n}{reduction\PYGZus{}error\PYGZus{}analysis}\PYG{p}{(}\PYG{n}{rd}\PYG{p}{,} \PYG{n}{d}\PYG{p}{,} \PYG{n}{rc}\PYG{p}{,}
                               \PYG{n}{error\PYGZus{}norms}\PYG{o}{=}\PYG{p}{[}\PYG{n}{induced\PYGZus{}norm}\PYG{p}{(}\PYG{n}{prod}\PYG{p}{)],} \PYG{n}{random\PYGZus{}seed}\PYG{o}{=}\PYG{l+m+mi}{77}\PYG{p}{)}
\PYG{k}{print}\PYG{p}{(}\PYG{n}{res}\PYG{p}{[}\PYG{l+s+s1}{\PYGZsq{}summary\PYGZsq{}}\PYG{p}{])}
\end{Verbatim}
\end{listing}

|VectorArrays| are ordered collections of vectors of same dimension, on which
standard linear algebra operations such as linear combination (|lincomb|) or
inner products (|dot|, |gramian|) can be performed. Selected degrees of
freedom can be extracted (|components|), as required for empirical interpolation.
All |VectorArray| methods can either act on single vectors or
the whole array.
Choosing arrays of vectors as elementary objects in \pyMOR (as opposed to single vectors) not only allows to
express many model reduction operations in a concise manner, it also allows
implementations which optimally exploit the vectorized structure of the
instructions for performance. An example of such an implementation is \pyMOR's
\mycode{NumpyVectorArray} (cf.\ Section~\ref{sec:benchmark}). Lists of single
vector objects inside external solvers can be easily managed using the
|ListVectorArray| class.

|Operators|, which in \pyMOR represent parametric matrices, bilinear forms, inner products,
functionals or nonlinear operators, can be applied to |VectorArrays|, yielding a new
|VectorArray| of results (|apply|). Access to (linear) solvers is provided via
|apply_inverse|, the Jacobian of an |Operator| is obtained as a new |Operator|
via the |jacobian| method.
Affinely decomposed (\ref{eq:affine_decomposition}) operators are
represented as |LincombOperators| which hold lists of
the |Operators| $B_q$ and the \mycode{ParameterFunctio}\-\mycode{nals} $\theta_q$.
|LincombOperators| can contain other |LincombOperators| as summands,
allowing an easy construction of arbitrarily nested affine decompositions,
which are automatically handled by \pyMOR's reduction algorithms.

Finally, |Discretizations| act as structured containers for |Operators|. For
instance, |StationaryDiscretization| has an |operator| and |rhs| attribute.
Moreover, |Dis|\-|cre|\-|ti|\-|za|\-|tions| can be solved for a given parameter, returning a
solution |VectorArray| (|solve|) which might be visualized using the |visualize|
method.
In |StationaryDiscretization|, |solve| for a given parameter |mu| is
implemented generically by calling
|self.operator.| |apply_inverse(self.rhs.as_vector(mu), mu=mu)|.
Other |Discretizations| might call optimized solution algorithms of an
external solver.

It is an important property of \pyMOR's interfaces that each method
either returns low-dimensional data or new \mycode{VectorArray},
\mycode{Operator} or \mycode{Discretization} objects. This ensures that
no high-dimensional data ever has to be communicated between the external solver
and \pyMOR and that no code for handling the solver-specific high-dimensional
data structures has to be added to \pyMOR.

Note that not only the high-dimensional model but also the reduced
low-di\-men\-sio\-nal model is represented by \mycode{VectorArrays},
\mycode{Operators} and \mycode{Discretizations}, implemented inside \pyMOR.
This allows to use all algorithms in \pyMOR with both high-
and low-dimensional objects. For instance, the reduced model could be interpreted
again as the high-dimensional model for an additional reduction step.

\begin{listing}[t]
  \caption{Methods used in Listing~\ref{alg:example_main}.}
  \label{alg:example_basic}
\begin{Verbatim}[commandchars=\\\{\}]
\PYG{k}{def} \PYG{n+nf}{pymor\PYGZus{}model}\PYG{p}{():}
    \PYG{n}{problem} \PYG{o}{=} \PYG{n}{ThermalBlockProblem}\PYG{p}{(}\PYG{n}{num\PYGZus{}blocks}\PYG{o}{=}\PYG{p}{(}\PYG{l+m+mi}{2}\PYG{p}{,} \PYG{l+m+mi}{2}\PYG{p}{))}
    \PYG{n}{d}\PYG{p}{,} \PYG{n}{\PYGZus{}} \PYG{o}{=} \PYG{n}{discretize\PYGZus{}elliptic\PYGZus{}cg}\PYG{p}{(}\PYG{n}{problem}\PYG{p}{,} \PYG{n}{diameter}\PYG{o}{=}\PYG{l+m+mf}{1.}\PYG{o}{/}\PYG{l+m+mf}{100.}\PYG{p}{)}
    \PYG{k}{return} \PYG{n}{d}\PYG{p}{,} \PYG{n}{d}\PYG{o}{.}\PYG{n}{h1\PYGZus{}0\PYGZus{}semi\PYGZus{}product}

\PYG{k}{def} \PYG{n+nf}{reduce\PYGZus{}naive}\PYG{p}{(}\PYG{n}{d}\PYG{p}{,} \PYG{n}{reductor}\PYG{p}{,} \PYG{n}{rb\PYGZus{}size}\PYG{p}{):}
    \PYG{n}{training\PYGZus{}set} \PYG{o}{=} \PYG{n}{d}\PYG{o}{.}\PYG{n}{parameter\PYGZus{}space}\PYG{o}{.}\PYG{n}{sample\PYGZus{}randomly}\PYG{p}{(}\PYG{n}{rb\PYGZus{}size}\PYG{p}{)}

    \PYG{n}{basis} \PYG{o}{=} \PYG{n}{d}\PYG{o}{.}\PYG{n}{operator}\PYG{o}{.}\PYG{n}{source}\PYG{o}{.}\PYG{n}{empty}\PYG{p}{()}
    \PYG{k}{for} \PYG{n}{mu} \PYG{o+ow}{in} \PYG{n}{training\PYGZus{}set}\PYG{p}{:}
        \PYG{n}{basis}\PYG{o}{.}\PYG{n}{append}\PYG{p}{(}\PYG{n}{d}\PYG{o}{.}\PYG{n}{solve}\PYG{p}{(}\PYG{n}{mu}\PYG{p}{))}

    \PYG{n}{rd}\PYG{p}{,} \PYG{n}{rc}\PYG{p}{,} \PYG{n}{\PYGZus{}} \PYG{o}{=} \PYG{n}{reductor}\PYG{p}{(}\PYG{n}{d}\PYG{p}{,} \PYG{n}{basis}\PYG{p}{)}
    \PYG{k}{return} \PYG{n}{rd}\PYG{p}{,} \PYG{n}{rc}
\end{Verbatim}
\end{listing}

\subsection{Implementation}\label{sec:python}

\pyMOR is implemented with \python, a managed, dynamically typed programming
language, which is easy to pick up (even for inexperienced programmers),
allowing quick and easy prototyping algorithms.

In contrast to \Matlab, \python does not have copy-on-write semantics
for assignment which, while allowing more precise control over data, often raises
the issue of object ownership. To alleviate the novice user from having to care too
much about ownership, \pyMOR enforces immutability on all \mycode{Discretizations} and
\mycode{Operators}. In combination with \python's dynamic memory management, this
makes the question of ownership irrelevant for these objects.
For \mycode{VectorArrays}, we have implemented shallow-copy/deep-copy-on-write
semantics.

While \python is designed as a general purpose language, the
\numpy \cite{Oliphant2007} package offers a multi-dimensional array class
with similar feature set and performance as \Matlab matrices. Additional
numerical algorithms and sparse matrix types can be found in the
\mycode{SciPy} \cite{JonesOliphantEtAl2001} package. Both packages are used extensively in \pyMOR for all
low-dimensional data structures as well as for \pyMOR's builtin discretization
toolbox.

\pyMOR's interfaces do not make any assumption on how the communication between
\pyMOR and the external solver is implemented by the interfacing classes, and
many communication patterns are conceivable, such as disk-based communication
via job and output files or network-based communication via a standard protocol
such as \mycode{xml-rpc} or custom protocols. Being a long-running general
purpose scripting language, \python is ideally suited for \pyMOR's
interface-based approach, offering a large selection of extension packages
for handling virtually any established input-output protocol.

However, in spirit of the tight coupling between \pyMOR and the external solver
as promoted by our design, we favor, whenever possible, to integrate the solver by
re-compiling it as a \python extension module, giving
\python direct access to the solver's data structures. This design
not only delivers maximum performance as no communication overhead is present, it
also allows to directly manipulate the solvers state from within
\python beyond the operations available via \pyMOR's interfaces. This allows
easy exploration of new ideas and offers the user an interactive \python debugging shell
with direct access to the solver's memory.
All external solvers in the following examples have been integrated with this
approach, showing its feasibility, even for \mycode{MPI}-distributed solvers running on
high-performance computing clusters. This approach has also been taken for the
integration of \pyMOR with the \mycode{BEST} battery simulation code as part of
the \textsc{Multibat} project \cite{OhlbergerRaveEtAl2014}.

\begin{listing}[t]
\caption{Further reduction methods in \texttt{example.py}.}
\label{alg:example_reduce_methods}
\begin{Verbatim}[commandchars=\\\{\}]
\PYG{k}{def} \PYG{n+nf}{reduce\PYGZus{}pod}\PYG{p}{(}\PYG{n}{d}\PYG{p}{,} \PYG{n}{reductor}\PYG{p}{,} \PYG{n}{product}\PYG{p}{,} \PYG{n}{snapshots}\PYG{p}{,} \PYG{n}{rb\PYGZus{}size}\PYG{p}{):}
    \PYG{n}{training\PYGZus{}set} \PYG{o}{=} \PYG{n}{d}\PYG{o}{.}\PYG{n}{parameter\PYGZus{}space}\PYG{o}{.}\PYG{n}{sample\PYGZus{}uniformly}\PYG{p}{(}\PYG{n}{snapshots}\PYG{p}{)}

    \PYG{n}{snapshots} \PYG{o}{=} \PYG{n}{d}\PYG{o}{.}\PYG{n}{operator}\PYG{o}{.}\PYG{n}{source}\PYG{o}{.}\PYG{n}{empty}\PYG{p}{()}
    \PYG{k}{for} \PYG{n}{mu} \PYG{o+ow}{in} \PYG{n}{training\PYGZus{}set}\PYG{p}{:}
        \PYG{n}{snapshots}\PYG{o}{.}\PYG{n}{append}\PYG{p}{(}\PYG{n}{d}\PYG{o}{.}\PYG{n}{solve}\PYG{p}{(}\PYG{n}{mu}\PYG{p}{))}
    \PYG{n}{basis}\PYG{p}{,} \PYG{n}{singular\PYGZus{}values} \PYG{o}{=} \PYG{n}{pod}\PYG{p}{(}\PYG{n}{snapshots}\PYG{p}{,} \PYG{n}{modes}\PYG{o}{=}\PYG{n}{rb\PYGZus{}size}\PYG{p}{,} \PYG{n}{product}\PYG{o}{=}\PYG{n}{product}\PYG{p}{)}

    \PYG{n}{rd}\PYG{p}{,} \PYG{n}{rc}\PYG{p}{,} \PYG{n}{\PYGZus{}} \PYG{o}{=} \PYG{n}{reductor}\PYG{p}{(}\PYG{n}{d}\PYG{p}{,} \PYG{n}{basis}\PYG{p}{)}
    \PYG{k}{return} \PYG{n}{rd}\PYG{p}{,} \PYG{n}{rc}

\PYG{k}{def} \PYG{n+nf}{reduce\PYGZus{}greedy}\PYG{p}{(}\PYG{n}{d}\PYG{p}{,} \PYG{n}{reductor}\PYG{p}{,} \PYG{n}{product}\PYG{p}{,} \PYG{n}{snapshots}\PYG{p}{,} \PYG{n}{rb\PYGZus{}size}\PYG{p}{):}
    \PYG{n}{training\PYGZus{}set} \PYG{o}{=} \PYG{n}{d}\PYG{o}{.}\PYG{n}{parameter\PYGZus{}space}\PYG{o}{.}\PYG{n}{sample\PYGZus{}uniformly}\PYG{p}{(}\PYG{n}{snapshots}\PYG{p}{)}
    \PYG{n}{ext\PYGZus{}alg} \PYG{o}{=} \PYG{n}{partial}\PYG{p}{(}\PYG{n}{gram\PYGZus{}schmidt\PYGZus{}basis\PYGZus{}extension}\PYG{p}{,} \PYG{n}{product}\PYG{o}{=}\PYG{n}{product}\PYG{p}{)}
    \PYG{n}{result} \PYG{o}{=} \PYG{n}{greedy}\PYG{p}{(}\PYG{n}{d}\PYG{p}{,} \PYG{n}{reductor}\PYG{p}{,} \PYG{n}{training\PYGZus{}set}\PYG{p}{,}
                    \PYG{n}{extension\PYGZus{}algorithm}\PYG{o}{=}\PYG{n}{ext\PYGZus{}alg}\PYG{p}{,} \PYG{n}{max\PYGZus{}extensions}\PYG{o}{=}\PYG{n}{rb\PYGZus{}size}\PYG{p}{,}
                    \PYG{n}{pool}\PYG{o}{=}\PYG{n}{new\PYGZus{}parallel\PYGZus{}pool}\PYG{p}{())}
    \PYG{k}{return} \PYG{n}{result}\PYG{p}{[}\PYG{l+s+s1}{\PYGZsq{}reduced\PYGZus{}discretization\PYGZsq{}}\PYG{p}{],} \PYG{n}{result}\PYG{p}{[}\PYG{l+s+s1}{\PYGZsq{}reconstructor\PYGZsq{}}\PYG{p}{]}

\PYG{k}{def} \PYG{n+nf}{reduce\PYGZus{}adaptive\PYGZus{}greedy}\PYG{p}{(}\PYG{n}{d}\PYG{p}{,} \PYG{n}{reductor}\PYG{p}{,} \PYG{n}{product}\PYG{p}{,} \PYG{n}{validation\PYGZus{}mus}\PYG{p}{,} \PYG{n}{rb\PYGZus{}size}\PYG{p}{):}
    \PYG{n}{ext\PYGZus{}alg} \PYG{o}{=} \PYG{n}{partial}\PYG{p}{(}\PYG{n}{gram\PYGZus{}schmidt\PYGZus{}basis\PYGZus{}extension}\PYG{p}{,} \PYG{n}{product}\PYG{o}{=}\PYG{n}{product}\PYG{p}{)}
    \PYG{n}{result} \PYG{o}{=} \PYG{n}{adaptive\PYGZus{}greedy}\PYG{p}{(}\PYG{n}{d}\PYG{p}{,} \PYG{n}{reductor}\PYG{p}{,} \PYG{n}{validation\PYGZus{}mus}\PYG{o}{=\PYGZhy{}}\PYG{n}{validation\PYGZus{}mus}\PYG{p}{,}
                             \PYG{n}{extension\PYGZus{}algorithm}\PYG{o}{=}\PYG{n}{ext\PYGZus{}alg}\PYG{p}{,} \PYG{n}{max\PYGZus{}extensions}\PYG{o}{=}\PYG{n}{rb\PYGZus{}size}\PYG{p}{,}
                             \PYG{n}{pool}\PYG{o}{=}\PYG{n}{new\PYGZus{}parallel\PYGZus{}pool}\PYG{p}{())}
    \PYG{k}{return} \PYG{n}{result}\PYG{p}{[}\PYG{l+s+s1}{\PYGZsq{}reduced\PYGZus{}discretization\PYGZsq{}}\PYG{p}{],} \PYG{n}{result}\PYG{p}{[}\PYG{l+s+s1}{\PYGZsq{}reconstructor\PYGZsq{}}\PYG{p}{]}
\end{Verbatim}
\end{listing}

\subsection{\pyMOR by example}\label{sec:example}
In the following, we consider a basic example of how four different high-dimensional
models of the form \eqref{eq:full_problem}, implemented with \pyMOR's discretization
toolkit, \mycode{FEniCS} \cite{LoggMardalEtAl2012}, \mycode{deal.II}
\cite{BangerthHartmannKanschat2007} and the \DUNE numerics environment \cite{BBDEKO08A,BBDEKO08B}, can be all reduced with
\pyMOR using identical reduction algorithms.

Listing~\ref{alg:example_main} contains the typical workflow in a \pyMOR
application. A |Discretization| object holding the high-dimensional
model is obtained first, here by by calling the |pymor_model| method.
In addition, |pymor_model| returns an |Operator| representing the inner product
on the space $V$.
Next, a |reductor| for performing the actual RB projection is selected.
Here, we choose the generic |reduce_stationary_coercive| method,
which will also assemble the error estimator \eqref{eq:estimator_efficiency}
according to \cite{BuhrEngwerEtAl2014}.
A |ParameterFunctional| which assigns to each $\mu \in \mathcal{P}$ the
coercivity estimate $\alpha_\mu$ is provided.
The reduced basis of size |10| and the resulting reduced model are then created
using the |reduce_naive| method, which returns the |Discretization| holding the
reduced model (|rd|) along with a |Reconstructor| object which is able to
perform the high-dimensional reconstruction \eqref{eq:reconstruction}.
Finally, the quality of the reduced model is evaluated using \pyMOR's
|reduction_error_analysis| method, which computes the model
reduction error and the error estimator effectivity for random parameters.

Listing~\ref{alg:example_basic} contains the |pymor_model| and |reduce_naive|
methods used in Listing~\ref{alg:example_main}. The former uses \pyMOR's
builtin discretization toolkit, first instantiating a problem description class
(line 2) and then discretizing the problem using first order continuous finite
elements (line 3). We consider here a classic $2 \times 2$ \lq thermal block\rq~test
problem of finding solutions for the stationary diffusion equation
\begin{equation}\label{eq:thermalblock}
\begin{aligned}
  - \nabla \cdot\big( a_\mu(x) \nabla u_\mu(x) \big) &= 1, \qquad x \in \Omega:=[0, 1]^2 \\
                                            u_\mu(x) &= 0, \qquad x \in \partial\Omega
\end{aligned}
\end{equation}
with diffusion coefficient $a_\mu$ given by the linear combination of indicator
functions
\begin{equation}\label{eq:thermalblock_diffusion}
  a_\mu(x) := \sum_{i=0}^m\sum_{j=0}^n \mu_{i,j}\cdot \chi_{[i/m, (i+1)/m] \times [j/n,
	  (j+1)/n]}(x),
\end{equation}
$m=n=2$, for parameters $\mu \in \mathcal{P} := [0.1, 1]^{2\times 2}$.
We can think of this as computing the heat distribution in a square material
composed of $2 \times 2$ blocks of different thermal conductivity $\mu_{i,j}$
while being uniformly heated and its temperature at the boundary kept constant at a value
of $0$.

The |reduce_naive| method simply computes the reduced basis by selecting a set
of random parameters (line 6) for which the high-dimensional solution is
computed and appended to the |basis| |VectorArray| (lines 7--9).

Listings~\ref{alg:example_main}~and \ref{alg:example_basic} already form a
complete \pyMOR application. However, to obtain good results, more advanced
basis generation techniques schould be used instead of |reduce_naive|.
Listing~\ref{alg:example_reduce_methods} contains three alternatives using
POD, the basic greedy algorithm (Algorithm~\ref{alg:greedy}) and an extended
adaptive version according to \cite{HaasdonkDihlmannEtAl2011}. |reduce_pod|
extends |reduce_naive| by computing a POD of the given solution snapshots (line
6). The |reduce_greedy| method mainly defers all work to \pyMOR's |greedy|
implementation (lines 12--14) which has to be called with a training set of parameters
(line 10) and a method for extending the existing reduced basis by the new
solution snapshot.
For this stationary problem, we can simply choose |gram_schmidt_basis_extension|
which orthonormalizes the new solution snapshot w.r.t.\ the old basis using a
stabilized Gram-Schmidt process including re-or\-tho\-nor\-maliza\-tion for
improved numerical accuracy.
Similarly, |reduce_adaptive_greedy| defers all work to \pyMOR's
|adaptive_greedy| method.
Both algorithms are provided a new default worker pool (|new_parallel_pool|) for
automatic parallelization of the reduction error estimation (line 1 of
Algorithm~\ref{alg:greedy}).

All four reduction methods are included in the
|example.py| script provided in the supplementary material, which contains
a slightly extended version of Listing~\ref{alg:example_main} as main function.
In addition to |pymor_model|, three further high-dimensional models are
available (cf.~Listing~\ref{alg:example_discretize_methods}):

|fenics_model| uses the \texttt{dolfin} \cite{LoggWells2010} \python module of the
\texttt{FEniCS} project to discretize a $4 \times 3$ \lq thermal block\rq\ problem
\eqref{eq:thermalblock}, \eqref{eq:thermalblock_diffusion} ($m=4, n=3$) using higher order
finite
\begin{listing}[H]
\caption{Further models in \texttt{example.py} (shortened).}
\label{alg:example_discretize_methods}
\begin{Verbatim}[commandchars=\\\{\}]
\PYG{k}{def} \PYG{n+nf}{fenics\PYGZus{}model}\PYG{p}{():}
    \PYG{o}{...}                             \PYG{c+c1}{\PYGZsh{} FEniCS code to setup discrete function space,}
    \PYG{n}{V}\PYG{p}{,} \PYG{n}{matrices}\PYG{p}{,} \PYG{n}{rhs}\PYG{p}{,} \PYG{n}{h1\PYGZus{}mat}  \PYG{o}{=} \PYG{o}{...} \PYG{c+c1}{\PYGZsh{} as well as system, inner product and rhs matrices}

    \PYG{k+kn}{from} \PYG{n+nn}{pymor.operators.fenics} \PYG{k+kn}{import} \PYG{n}{FenicsMatrixOperator}
    \PYG{k+kn}{from} \PYG{n+nn}{pymor.vectorarrays.fenics} \PYG{k+kn}{import} \PYG{n}{FenicsVector}
    \PYG{n}{coeffs} \PYG{o}{=} \PYG{p}{[}\PYG{n}{ProjectionParameterFunctional}\PYG{p}{(}\PYG{l+s+s1}{\PYGZsq{}diffusion\PYGZsq{}}\PYG{p}{,} \PYG{p}{(}\PYG{l+m+mi}{4}\PYG{p}{,} \PYG{l+m+mi}{3}\PYG{p}{),} \PYG{p}{(}\PYG{l+m+mi}{3} \PYG{o}{\PYGZhy{}} \PYG{n}{y} \PYG{o}{\PYGZhy{}} \PYG{l+m+mi}{1}\PYG{p}{,} \PYG{n}{x}\PYG{p}{))}
              \PYG{k}{for} \PYG{n}{x} \PYG{o+ow}{in} \PYG{n+nb}{range}\PYG{p}{(}\PYG{l+m+mi}{4}\PYG{p}{)} \PYG{k}{for} \PYG{n}{y} \PYG{o+ow}{in} \PYG{n+nb}{range}\PYG{p}{(}\PYG{l+m+mi}{3}\PYG{p}{)]}
    \PYG{n}{ops} \PYG{o}{=} \PYG{p}{[}\PYG{n}{FenicsMatrixOperator}\PYG{p}{(}\PYG{n}{m}\PYG{p}{,} \PYG{n}{V}\PYG{p}{,} \PYG{n}{V}\PYG{p}{)} \PYG{k}{for} \PYG{n}{m} \PYG{o+ow}{in} \PYG{n}{matrices}\PYG{p}{]}
    \PYG{n}{op} \PYG{o}{=} \PYG{n}{LincombOperator}\PYG{p}{(}\PYG{n}{ops}\PYG{p}{,} \PYG{n}{coeffs}\PYG{p}{)}
    \PYG{n}{rhs} \PYG{o}{=} \PYG{n}{VectorFunctional}\PYG{p}{(}\PYG{n}{ListVectorArray}\PYG{p}{([}\PYG{n}{FenicsVector}\PYG{p}{(}\PYG{n}{rhs}\PYG{p}{,} \PYG{n}{V}\PYG{p}{)]))}
    \PYG{n}{h1\PYGZus{}product} \PYG{o}{=} \PYG{n}{FenicsMatrixOperator}\PYG{p}{(}\PYG{n}{h1\PYGZus{}mat}\PYG{p}{,} \PYG{n}{V}\PYG{p}{,} \PYG{n}{V}\PYG{p}{,} \PYG{n}{name}\PYG{o}{=}\PYG{l+s+s1}{\PYGZsq{}h1\PYGZus{}0\PYGZus{}semi\PYGZsq{}}\PYG{p}{)}

    \PYG{n}{param\PYGZus{}space} \PYG{o}{=} \PYG{n}{CubicParameterSpace}\PYG{p}{(}\PYG{n}{op}\PYG{o}{.}\PYG{n}{parameter\PYGZus{}type}\PYG{p}{,} \PYG{l+m+mf}{0.1}\PYG{p}{,} \PYG{l+m+mf}{1.}\PYG{p}{)}
    \PYG{n}{d} \PYG{o}{=} \PYG{n}{StationaryDiscretization}\PYG{p}{(}\PYG{n}{op}\PYG{p}{,} \PYG{n}{rhs}\PYG{p}{,} \PYG{n}{products}\PYG{o}{=}\PYG{p}{\PYGZob{}}\PYG{l+s+s1}{\PYGZsq{}h1\PYGZus{}0\PYGZus{}semi\PYGZsq{}}\PYG{p}{:} \PYG{n}{h1\PYGZus{}product}\PYG{p}{\PYGZcb{},}
                                 \PYG{n}{parameter\PYGZus{}space}\PYG{o}{=}\PYG{n}{param\PYGZus{}space}\PYG{p}{)}
    \PYG{k}{return} \PYG{n}{d}\PYG{p}{,} \PYG{n}{d}\PYG{o}{.}\PYG{n}{h1\PYGZus{}0\PYGZus{}semi\PYGZus{}product}

\PYG{k}{def} \PYG{n+nf}{dealii\PYGZus{}model}\PYG{p}{():}
    \PYG{k+kn}{from} \PYG{n+nn}{dealii\PYGZus{}example} \PYG{k+kn}{import} \PYG{n}{ElasticityExample}
    \PYG{n}{example} \PYG{o}{=} \PYG{n}{ElasticityExample}\PYG{p}{(}\PYG{n}{refine\PYGZus{}steps}\PYG{o}{=}\PYG{l+m+mi}{9}\PYG{p}{)}

    \PYG{k+kn}{from} \PYG{n+nn}{pydealii.pymor.operator} \PYG{k+kn}{import} \PYG{n}{DealIIMatrixOperator}
    \PYG{k+kn}{from} \PYG{n+nn}{pydealii.pymor.vectorarray} \PYG{k+kn}{import} \PYG{n}{DealIIVector}
    \PYG{n}{coeffs} \PYG{o}{=} \PYG{p}{[}\PYG{n}{ProjectionParameterFunctional}\PYG{p}{(}\PYG{l+s+s1}{\PYGZsq{}mu\PYGZsq{}}\PYG{p}{,} \PYG{n+nb}{tuple}\PYG{p}{()),}
              \PYG{n}{ProjectionParameterFunctional}\PYG{p}{(}\PYG{l+s+s1}{\PYGZsq{}lambda\PYGZsq{}}\PYG{p}{,} \PYG{n+nb}{tuple}\PYG{p}{())]}
    \PYG{n}{ops} \PYG{o}{=} \PYG{p}{[}\PYG{n}{DealIIMatrixOperator}\PYG{p}{(}\PYG{n}{example}\PYG{o}{.}\PYG{n}{mu\PYGZus{}mat}\PYG{p}{()),}
           \PYG{n}{DealIIMatrixOperator}\PYG{p}{(}\PYG{n}{example}\PYG{o}{.}\PYG{n}{lambda\PYGZus{}mat}\PYG{p}{())]}
    \PYG{n}{op} \PYG{o}{=} \PYG{n}{LincombOperator}\PYG{p}{(}\PYG{n}{ops}\PYG{p}{,} \PYG{n}{coeffs}\PYG{p}{)}
    \PYG{n}{rhs} \PYG{o}{=} \PYG{n}{VectorFunctional}\PYG{p}{(}\PYG{n}{ListVectorArray}\PYG{p}{([}\PYG{n}{DealIIVector}\PYG{p}{(}\PYG{n}{example}\PYG{o}{.}\PYG{n}{rhs}\PYG{p}{())]))}

    \PYG{n}{param\PYGZus{}space} \PYG{o}{=} \PYG{n}{CubicParameterSpace}\PYG{p}{(}\PYG{n}{op}\PYG{o}{.}\PYG{n}{parameter\PYGZus{}type}\PYG{p}{,} \PYG{l+m+mf}{1.}\PYG{p}{,} \PYG{l+m+mf}{10.}\PYG{p}{))}
    \PYG{n}{d} \PYG{o}{=} \PYG{n}{StationaryDiscretization}\PYG{p}{(}\PYG{n}{op}\PYG{p}{,} \PYG{n}{rhs}\PYG{p}{,} \PYG{n}{products}\PYG{o}{=}\PYG{p}{\PYGZob{}}\PYG{l+s+s2}{\PYGZdq{}energy\PYGZdq{}}\PYG{p}{:} \PYG{n}{ops}\PYG{p}{[}\PYG{l+m+mi}{0}\PYG{p}{]\PYGZcb{},}
                                 \PYG{n}{parameter\PYGZus{}space}\PYG{o}{=}\PYG{n}{param\PYGZus{}space}\PYG{p}{)}
    \PYG{k}{return} \PYG{n}{d}\PYG{p}{,} \PYG{n}{d}\PYG{o}{.}\PYG{n}{energy\PYGZus{}product}

\PYG{k}{def} \PYG{n+nf}{dune\PYGZus{}model}\PYG{p}{():}
    \PYG{k+kn}{from} \PYG{n+nn}{spe10} \PYG{k+kn}{import} \PYG{n}{examples}\PYG{p}{,} \PYG{n}{wrapper}
    \PYG{n}{example} \PYG{o}{=} \PYG{n}{examples}\PYG{p}{[}\PYG{l+s+s1}{\PYGZsq{}aluconformgrid\PYGZsq{}}\PYG{p}{][}\PYG{l+s+s1}{\PYGZsq{}pdelab\PYGZsq{}}\PYG{p}{][}\PYG{l+s+s1}{\PYGZsq{}istl\PYGZsq{}}\PYG{p}{](}\PYG{l+s+s1}{\PYGZsq{}[60 220 85]\PYGZsq{}}\PYG{p}{,} \PYG{n+nb+bp}{True}\PYG{p}{)}
    \PYG{n}{d} \PYG{o}{=} \PYG{n}{wrapper}\PYG{p}{[}\PYG{n}{example}\PYG{o}{.}\PYG{n}{discretization}\PYG{p}{()]}

    \PYG{n}{param\PYGZus{}ranges} \PYG{o}{=} \PYG{p}{\PYGZob{}}\PYG{l+s+s1}{\PYGZsq{}blockade\PYGZsq{}}\PYG{p}{:} \PYG{p}{(}\PYG{l+m+mf}{1e\PYGZhy{}4}\PYG{p}{,} \PYG{l+m+mi}{1}\PYG{p}{),} \PYG{l+s+s1}{\PYGZsq{}sink\PYGZsq{}}\PYG{p}{:} \PYG{p}{(}\PYG{l+m+mi}{0}\PYG{p}{,} \PYG{l+m+mi}{1}\PYG{p}{)\PYGZcb{}}
    \PYG{n}{param\PYGZus{}space} \PYG{o}{=} \PYG{n}{CubicParameterSpace}\PYG{p}{(}\PYG{n}{op}\PYG{o}{.}\PYG{n}{parameter\PYGZus{}type}\PYG{p}{,} \PYG{n}{ranges}\PYG{o}{=}\PYG{n}{param\PYGZus{}ranges}\PYG{p}{)}
    \PYG{n}{d} \PYG{o}{=} \PYG{n}{d}\PYG{o}{.}\PYG{n}{with\PYGZus{}}\PYG{p}{(}\PYG{n}{parameter\PYGZus{}space}\PYG{o}{=}\PYG{n}{param\PYGZus{}space}\PYG{p}{)}
    \PYG{k}{return} \PYG{n}{d}\PYG{p}{,} \PYG{n}{d}\PYG{o}{.}\PYG{n}{energy\PYGZus{}0\PYGZus{}product}\PYG{o}{.}\PYG{n}{assemble}\PYG{p}{(}\PYG{l+m+mf}{1e\PYGZhy{}4}\PYG{p}{)}
\end{Verbatim}
\end{listing}%
\UndefineShortVerb{\|}
\begin{figure}[t]
  \footnotesize%
  \centering%
  \includegraphics[clip, trim=0 636 263 0]{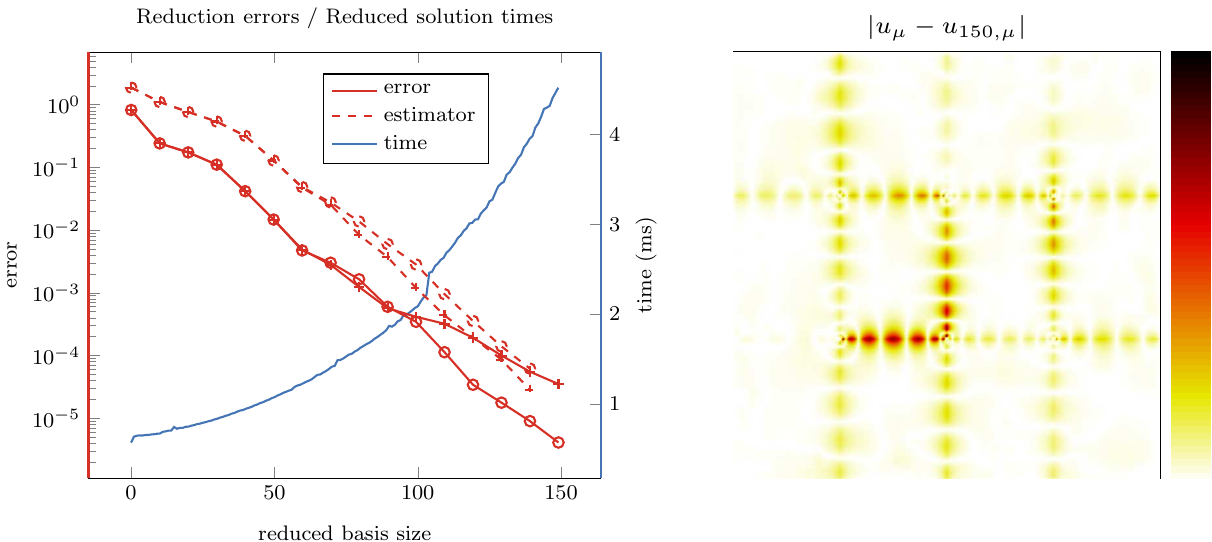}%
  \caption{%
  Error analysis of the RB approximation of
    \texttt{fenics\textunderscore model} using \texttt{reduce\textunderscore greedy}.
    \emph{Left:} Maximum $H^1$-model order reduction error (red, solid) on a test set of 1000 parameters and maximum
    estimated error (red, dashed) on the training set in dependence on the reduced
    basis ($\mathcal{S}_{train,1} := \{0.1,
    1\}^{4\times 3}$: \protect\includegraphics[clip, trim=0 785 605 0]{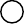}, $\mathcal{S}_{train, 2}:=\{0.1, 0.55, 1\}^{4\times 3}$: $+$);
    average time for the computation of a reduced solution including error estimation in dependence on the reduced basis size (blue).
    \emph{Right:} Plot of the absolute difference (dark: $0$, light: $6.03\cdot 10^{-7}$)
    between the detailed and the reduced solution $|u_\mu - u_{150, \mu}|$ (for $\mathcal{S}_{train,2}$) for the worst approximated parameter
    of the test set.
  }
  \label{fig:thermalblock}
\end{figure}
\DefineShortVerb{\|}
\noindent elements.
The resulting matrix objects are then wrapped by \pyMOR's
|Fenics|\-|Matrix|\-|Operator| and |FenicsVector| classes
which translate \pyMOR interface calls into appropriate operations on the
underlying \texttt{FEniCS} data structures (lines 8, 10, 11).
The affine decomposition \eqref{eq:affine_decomposition} of the problem is encoded
by defining appropriate |ParameterFunctionals| and then
constructing an operator representing the decomposition (lines 6, 7, 9).
Finally, an appropriate parameter space is chosen, and everything is wrapped up
in a generic |StationaryDiscretizaion| object (lines 12--14).
Figure~\ref{fig:thermalblock} (left) shows model reduction errors and timings for
|fenics_model| (second order finite elements, 361.201 degrees of freedom)
reduced by |reduce_greedy| using the training sets
$\mathcal{S}_{train,1}:=\{0.1, 1\}^{4\times 3}$ and $\mathcal{S}_{train,2}:=\{0.1, 0.55, 1\}^{4\times 3}$.
We observe that choosing a too small training set leads to
\emph{overfitting}: while the solution is very well approximated
for parameters in the training set, the approximation quality is not
maintained for parameters not contained in the training set.
\UndefineShortVerb{\|}%
The error $|u_\mu - u_{N, \mu}|$ for the final basis size
($\mathcal{S}_{train,2}$) is shown in Figure~\ref{fig:thermalblock} (right).
\DefineShortVerb{\|}

\UndefineShortVerb{\|}
\newcommand{\DEALHEIGHT}{0.84\linewidth}
\newcommand{\DEALWIDTH}{1.13\linewidth}
\newcommand{\DEALEPSSCALE}{0.92\textwidth}
\def\imagebox#1{\vtop to 45mm{\null\hbox{#1}\vfill}}

\begin{figure}[t]
\centering
  \begin{subfigure}[b]{.33\linewidth}
      \imagebox{\includegraphics[keepaspectratio, width=\DEALEPSSCALE]{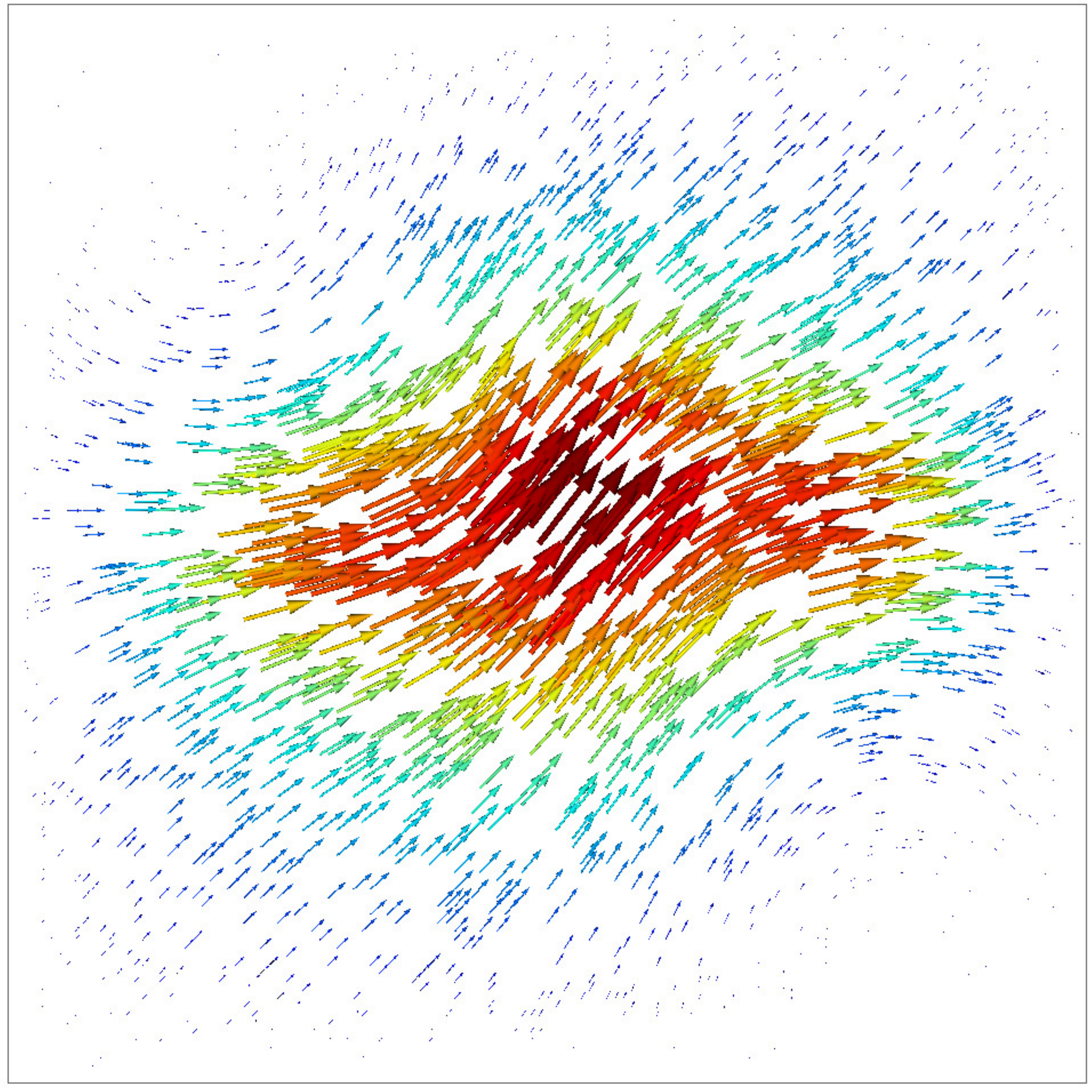}}
  \end{subfigure}~
  \begin{subfigure}[b]{.33\linewidth}
      \imagebox{\includegraphics[keepaspectratio,width=\DEALEPSSCALE]{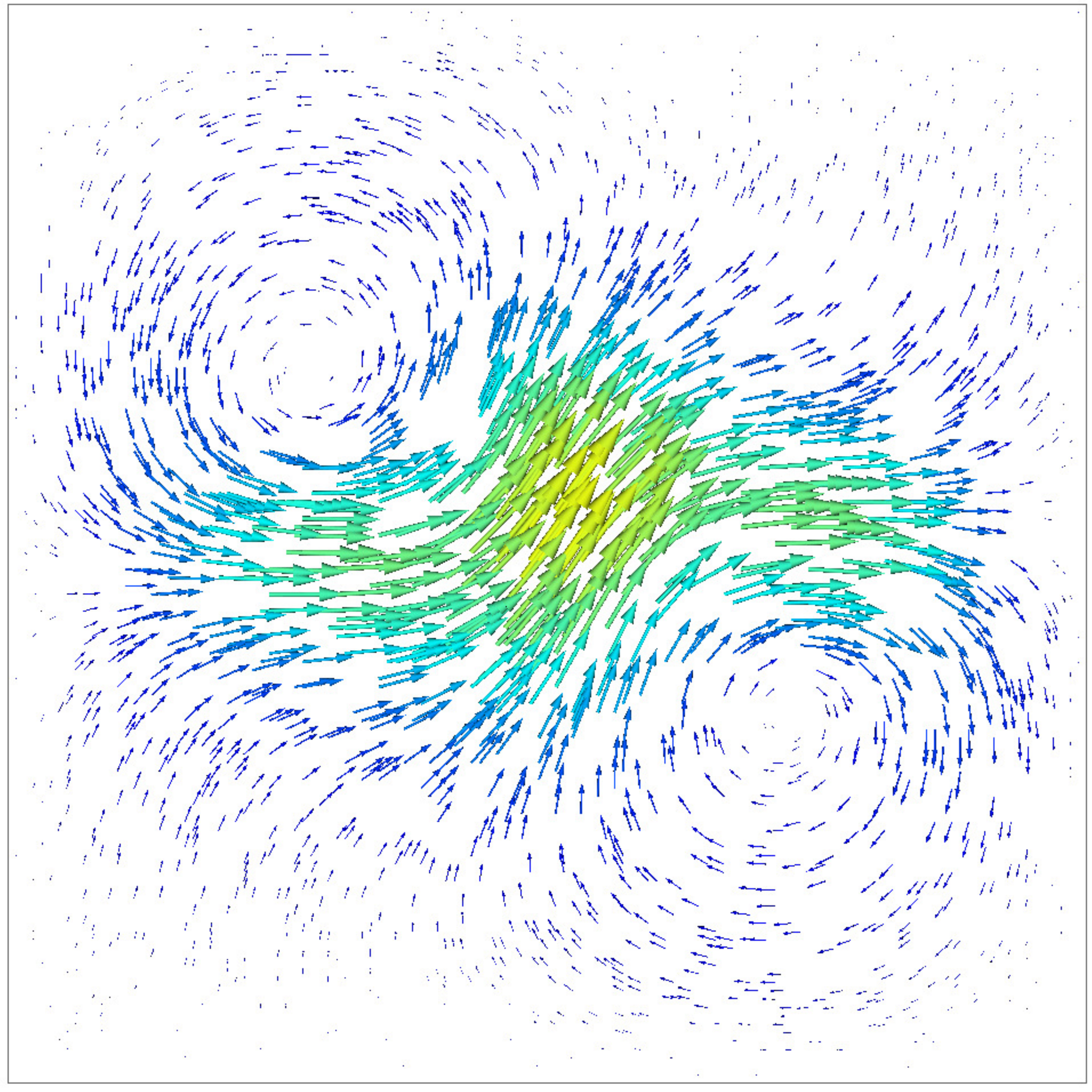}}
  \end{subfigure}~~
  \begin{subfigure}[b]{.33\linewidth}\imagebox{%
  \includegraphics{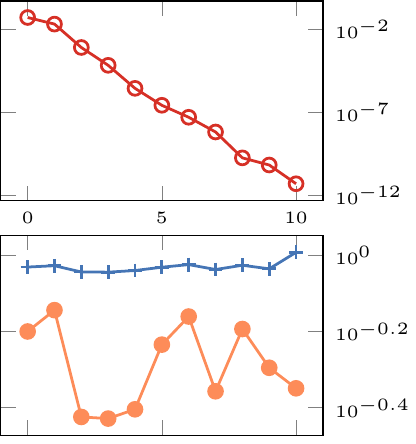}}
  \end{subfigure}
 \caption{Sample solutions and error analysis of the RB approximation of \mycode{dealii\_model} using \mycode{adaptive\_greedy}.
          Plot of displacement field $u_{\mu,\lambda}$ for $(\mu, \lambda)=(1, 1)$ (\textit{left}) and
                $(\mu, \lambda)=(1, 10)$ (\textit{middle}) with color representing magnitude (blue $=0$, red $=0.033$).
          \textit{Right}: Plot of maximum energy model reduction error (\textit{top}, \protect\includegraphics[clip, trim=0 785 605 0]{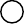})
                and error estimator effectivity (\textit{bottom}, min: \protect\includegraphics[clip, trim=0 785 605 0]{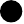}, max: $+$) over a test set of 10 random parameters
                against the size of the reduced basis ($\left[0,10\right]$).}
\label{fig::dealii}
\end{figure}
\DefineShortVerb{\|}

\UndefineShortVerb{\|}
\begin{figure}[t]
  \footnotesize%
  \centering%
  \includegraphics[clip, trim=0 672 247 0]{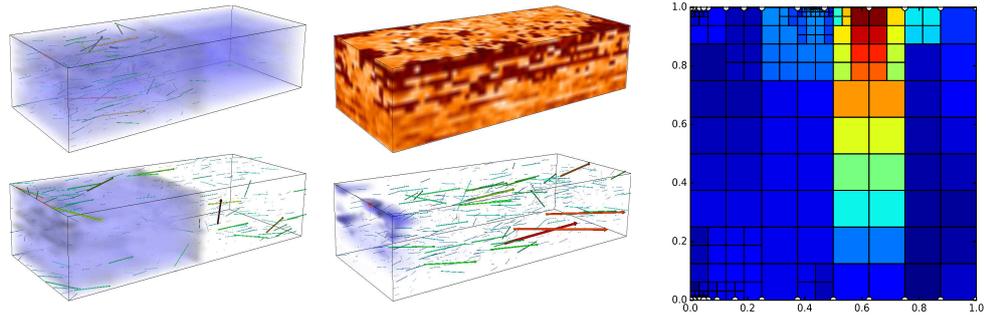}
  \caption{%
    Data function, sample solutions and adaptively refined set of training samples of \mycode{dune\_model} using \mycode{adaptive\_greedy}.
    \emph{Top center:} logarithmic plot of the Frobenius norm of the anisotropic SPE10 model2 permeability tensor (dark: $6.65\cdot 10^{-8}$, light: $2\cdot 10^5$).
    \emph{Left and bottom center:} volume plot of the pressure $u_\mu$ and vector plot of the reconstructed Darcy velocity $-\kappa_\mu \nabla u_\mu$ (colored and scaled by magnitude) for several parameters (blue: weak, red: strong): $\mu = (10^{-4}, 0)$ bottom left, $\mu = (10^{-4}, 1)$ top left and $\mu = (1, 0)$ bottom center.
    The first parameter component models the existence of a blockade  in the middle (enabled: $10^{-4}$, disabled: $1$) and the second parameter component acts as a switch for a sink near the blockade.
    \emph{Right:} plot of the adaptively refined training set for a reduced basis of size 50 (colored contribution of the local error indicators, blue: low, red: high) and selected training parameters (white circles).
    Note the strong influence of the first parameter component (which influences the operator), as opposed to the second parameter component (which only influences the right hand side).
  }
  \label{fig:dune}
\end{figure}
\DefineShortVerb{\|}

In |discretize_dealii|, we consider the two-dimensional linear elasticity problem
\begin{equation}\label{eq:lin_elast}
  \begin{aligned}
  -\nabla\lambda\left(\nabla \cdot u_{\mu, \lambda}\right)-\left(\nabla\cdot \mu\nabla\right)u_{\mu, \lambda} -\nabla\cdot\mu\left(\nabla u_{\mu, \lambda}\right)^T&=f
    \quad \textnormal{in} \,\,\Omega=[0,1]^2
  \end{aligned}
\end{equation}
with Lam\'e parameters $\mu,\lambda \in [1, 10]$ and homogeneous Dirichlet boundary conditions
from the tutorial documentation for the \texttt{deal.II}\footnote{\url{https://dealii.org/8.1.0/doxygen/deal.II/step_8.html}}
\cite{BangerthHartmannKanschat2007} \Cpp{}\ solver library (cf.~Figure \ref{fig::dealii}, left, middle).
The tutorial program was refactored into the |ElasticityExample| class and \python
bindings for this class were added (line 17).
Moreover, we implemented \python bindings for the
\texttt{deal.II} |SparseMatrix| and |Vector| classes, as well as corresponding
\pyMOR wrapper classes. These are used to make the \texttt{deal.II} matrices and
right-hand side vector provided by |ElasticExample| available as \pyMOR
|Operators| (lines 23, 24, 26).
The |ElasticityExample| calculates a solution for the displacement field $u_{\mu, \lambda}$ by discretizing
\eqref{eq:lin_elast} with first order continuous Galerkin finite elements ($65.536$ degrees of freedom) and solves
the resulting linear system with a conjugate gradient method.
The \python bindings and \pyMOR wrappers are available in the
\texttt{pymor-deal.II} package. Figure \ref{fig::dealii} (right) shows model reduction errors and estimator
effectivities for \mycode{dealii\_model} reduced by \mycode{adaptive\_greedy} for a reduced basis of size 10.

|dune_model| uses the \DUNE numerics environment to discretize a multi-scale single phase flow problem with the highly heterogeneous and anisotropic SPE10 model2 permeability field\footnote{\url{http://www.spe.org/web/csp/datasets/set02.htm}}, with inflow boundary conditions at $x_1 = 0$, a fixed pressure at $x_1 = 5$ and no flow at the other boundaries:
\begin{equation}\label{eq:spe10}
\begin{aligned}
  - \nabla \cdot\big( \kappa_\mu \nabla u_\mu \big) &= f_\mu, \qquad \text{in } \Omega:=[0, 2] \times [0, 5] \times [0, 1].
\end{aligned}
\end{equation}
The parameter $\mu$ allows to toggle the presence of a blockade at $x_1 = 2.5$ and a sink near the blockade (compare Figure~\ref{fig:dune}, left and middle).
Problem \eqref{eq:spe10} is discretized with \dune{gdt} \cite{dunegdt} using first order continuous finite elements on a tetrahedral grid with $6.7\cdot 10^6$ elements ($1.2\cdot 10^6$ degrees of freedom).
The integration with \pyMOR is based on \dune{pymor} which automatically wraps the resulting
discretization objects defined in the \dune{hdd} module as \pyMOR
|Discretizations| (line 34) taking the affine decomposition of the problem into account.
Only the parameter space is additionally chosen (lines 36--37).
Figure~\ref{fig:dune} shows the adaptively refined training set and selected training parameters for a reduced basis of size 50, which yields a maximum absolute model reduction energy error of $5.5 \cdot 10^{-10}$ over a set of 100 randomly chosen test parameters.

Again, let us note that all four high-dimensional models can be reduced with the
exact same model reduction code, ranging from the trivial |reduce_naive|
algorithm to the sophisticated adaptive snashpot selection in
|reduce_adaptive_greedy|.
In fact, the |example.py| script provided in the supplementary material allows to choose any of
the 16 possible combinations between PDE solver and reduction method.

\UndefineShortVerb{\|}

\subsection{Parallelism in \pyMOR}\label{sec:parallel}

\DefineShortVerb{\|}
In many application areas of reduced order modeling, not only the efficiency of
the reduced model but also the required time for generating the model needs to
be take into account.
Thus, RB software should be able
to perform offline computations with good computational efficiency. For modern
computing architectures this requires parallelization of algorithms.

In a greedy basis generation algorithm (Algorithm \ref{alg:greedy}),
the main computational work is made up by three types of operations:
1.~reduction error estimation on $\mathcal{S}_{train}$,
2.~computation of the solution
snapshot $u_{\mu^*}$, 3.~reduced basis extension and reduction of the high-dimensional model.

For the parallelization of the high-dimensional operations in steps 2 and 3, \pyMOR relies on already existing,
high-performance parallelization of the external solver. Since \pyMOR's
interfaces require no communication of any high-dimensional data and are
completely implementation agnostic, memory distributed vector data can be handled
via the \mycode{VectorArray} interface as efficiently as any non-distributed data.

To ease the integration of \mycode{MPI} distributed codes, \pyMOR
offers tools which allow to automatically support \mycode{MPI} parallel
use of the solver when \pyMOR bindings for the sequential case already exist.
For instance, to parallelize |fenics_model| in
Listing~\ref{alg:example_discretize_methods}, one would simply execute:
\begin{code}
d = mpi_wrap_discretization(lambda: fenics_model()[0],
                            use_with=True, pickle_subtypes=False)
prod = d.h1_0_semi_product
\end{code}
When the script is executed with |mpirun|, an event loop is
launched on all |MPI| ranks except for rank 0 which executes the main script.
|mpi_wrap_discretization| then instructs each rank to execute the function
given as first argument to obtain a local |Discretization| object.
The |Discretization| returned by |mpi_wrap_|-|discretization| and all contained
|Operators| will then use the event loop to issue |MPI| distributed operations on the
rank-local objects when corresponding interface methods are called.
In Section~\ref{sec:burgers}, we consider another example where \pyMOR's |MPI|
wrappers are used.

For the parallelization of step 1 and similar embarrassingly parallel tasks
which require little to no communication, \pyMOR provides an abstraction
layer for existing \python parallelization solutions based on a simple
worker pool concept (\mycode{pymor.} \mycode{parallel}).
For instance, line 1 in Algorithm~\ref{alg:greedy} could be parallelized
by executing
\begin{code}
numpy.argmax(pool.map(lambda mu,rd: rd.estimate(rd.solve(mu), mu),
                      training_set, rd=reduced_discretization))
\end{code}
The function and all arguments are automatically serialized and distributed to
the workers of the pool, ensuring that immutable data is only
communicated once.

\pyMOR currently provides a worker pool implementation based on the
\mycode{IPython} \cite{PerezGranger2007} toolkit, which allows easy parallel
computation with large collections of heterogeneous compute nodes, and an
\mycode{MPI}-based implementation using \pyMOR's event loop which can
seamlessly be used in conjunction with external solvers using the same
event loop.
The \mycode{MPI}-based worker pool was also used for the experiment in
Figure~\ref{fig:thermalblock}, employing 16 cores of the compute server
described in Section~\ref{sec:benchmark}.
For the larger training set $\mathcal{S}_{train,2}$ with 531.441 parameters,
the total offline computation time was 11 hours, of which 5 hours were
spent on error estimation.
Performing the error estimation without parallelization would have required
an additional 37 hours of computation time.

\UndefineShortVerb{\|}

\section{Performance evaluation}\label{sec:experiments}

In this section we evaluate the applicability and performance of \pyMOR's design
approach by considering technical benchmarks of some of \pyMOR's interfaces, as
well as a challenging nonlinear large-scale model order reduction
problem.

\subsection{Benchmarks}\label{sec:benchmark}

The main goal of this section is to compare the performance of \pyMOR's
\mycode{VectorArray} and \mycode{Operator} interfaces when used to access
external high-di\-men\-sio\-nal solver data structures to native
implementations of these classes. Moreover, we want to investigate
possible performance benefits for vectorized \mycode{VectorArray}
implementations.

As native implementations within \pyMOR we consider \mycode{NumpyVectorArray},
which allows vectorized operations on vectors by internally holding an
appropriately sized two-dimensional \numpy array, as well as \mycode{ListVectorArray}
maintaining a \python list data structure holding vector objects implemented as
one-dimensional \numpy arrays. The inner
product in the \mycode{pod} benchmark is
implemented with \mycode{NumpyMatrixOperators} holding sparse \mycode{SciPy}
matrices coming from \pyMOR's own discretization toolbox.

The external solver code is based on the \DUNE numerics environment \cite{BBDEKO08A,BBDEKO08B},
centered around the discretization toolbox \dune{gdt} \cite{dunegdt} (compare Section \ref{sec:example}) and is compiled as a \python extension module as
described in Section~\ref{sec:python}.

Our benchmarks were executed on a dual socket compute server equipped with
two Intel Xeon
E5-2698 v3 CPUs with 16 cores running at 2.30GHz each and 256GB of memory
available. All benchmarks were performed as single-threaded processes.

\paragraph{Vector array benchmark}
We consider the \mycode{axpy} method of the \mycode{Vector}\-\mycode{Array}\-\mycode{Interface},
which performs a vectorized BLAS-conforming \mycode{axpy}
operation, i.e.\ pairwise in-place addition of the vectors in the array with vectors
of a second array multiplied by a scalar factor.

\begin{figure}[t]
  \footnotesize%
  \centering%
  \includegraphics[clip, trim=0 636 247 0]{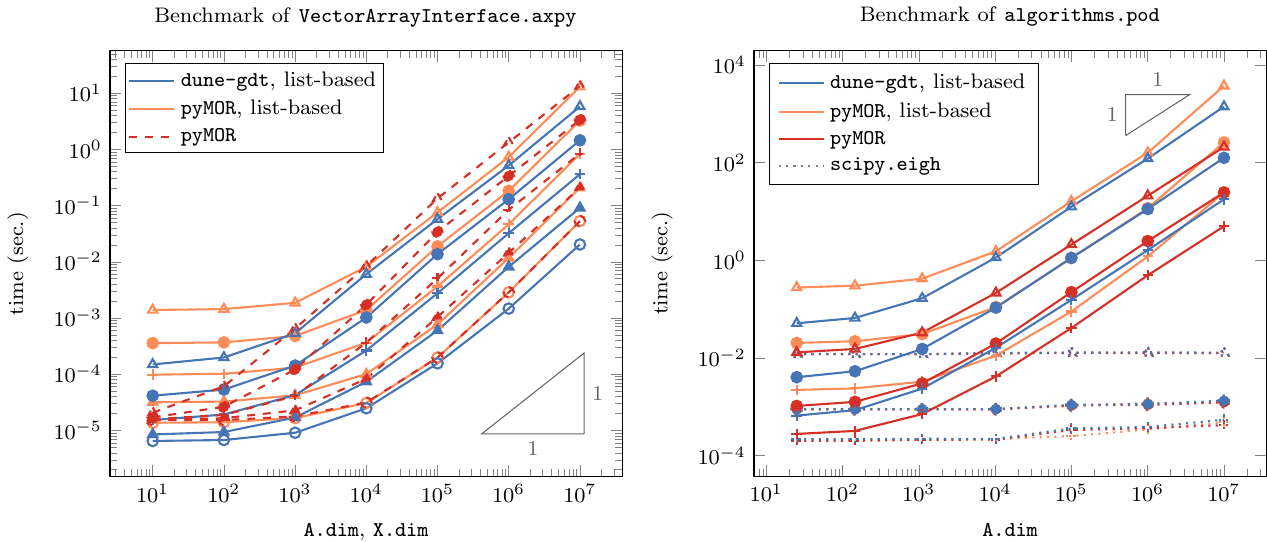}%
  \caption{%
    Log/log plot of the measured execution time of \mycode{A.axpy} (left) and the POD algorithm (right) for different implementations and several lengths of the vector array (\mycode{len(A)==1}: \protect\includegraphics[clip, trim=0 785 605 0]{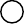}, \mycode{len(A)==4}: $\blacktriangle$, \mycode{len(A)==16}: $+$, \mycode{len(A)==64}: \protect\includegraphics[clip, trim=0 785 605 0]{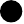}, \mycode{len(A)==256}: $\triangle$).
    \emph{Left:} Comparison of \mycode{A.axpy(X)} with \mycode{len(X)==len(A)} for several implementations of \mycode{A}.
    \emph{Right:} Comparison of \mycode{pod(A=A, modes=10, product==h1\_0, orthonormalize=False, check=False)} for several implementations of \mycode{A} and the \mycode{h1\_0} product (solid) and the respective time spent in \mycode{scipy.eigh} (dotted).
  }
  \label{figure::benchmarks}
\end{figure}

As we observe in Figure
\ref{figure::benchmarks} (left), both
\mycode{List}\-\mycode{Vector}\-\mycode{Array}-based implementations show about the same performance
for sufficiently large array dimensions. In fact, the \DUNE-based implementation
(\dune{gdt}, list based) actually performs better than the \numpy-based
implementation (\pyMOR, list based), showing the tight
integration between \pyMOR and the external solver.
The \mycode{Numpy}\-\mycode{Vector}\-\mycode{Array} implementation, on the other hand,
cannot benefit from vectorization for larger array dimensions.\footnote{We assume that this is due to the fact that
\numpy offers no native \mycode{axpy} operations, such that a large temporary array
has to be created holding all to be added and scaled vectors. It is
	planned to improve performance of \mycode{NumpyVectorArray.axpy} by
	directly calling out to BLAS (cf.\
	\url{https://github.com/pymor/pymor/issues/73}).}

\paragraph{POD benchmark}
As detailed previously (e.g., Section~\ref{sec:high_dim_operations}), the Gram-Schmidt and POD
algorithms are important tools in the context of model reduction.
The implementation of a numerically stable Gram-Schmidt or POD algorithm might
not be completely straightforward and it is an important benefit that \pyMOR's
interface design allows to provide tried and tested implementations of these
algorithms which can automatically be used with any external solver integrated
with \pyMOR.

\pyMOR's POD algorithm mainly consist of three steps with different complexities:
1.~computation of a Gramian matrix with respect to the given inner product
(\mycode{product.} \mycode{apply2}), 2.~computation of the eigenvalue decomposition of the
Gramian (using the \mycode{SciPy} \mycode{eigh} method) and 3. mapping
right-singular vectors to left-singular vectors (by calling \mycode{lincomb} on
the original \mycode{VectorArray}). Note that the computational cost for steps 1
and 3 depends on the \mycode{VectorArray} and \mycode{Operator} implementation, scaling linearly with
the array dimension and quadratically (resp.\ linearly) with the array length.
The computational cost for step 2 is independent of the space dimension,
and only increases with the number of given vectors.

As the inner product for both benchmarks and all implementations we have chosen the full
$H^1$-product matrix stemming from a first order continuous finite element discretization
over the same structured triangular grid on the unit square.

As we observe again in  Figure~\ref{figure::benchmarks} (right) both
\mycode{ListVectorArray}-based implementation show roughly equal performance.
However, the vectorized \pyMOR implementation is able to clearly outperform
both other implementations thanks to the fact that the computationally dominant steps 1 and 3 of
the algorithm can be expressed idiomatically via a single interface
call.
This shows that
\mycode{VectorArray} implementations can indeed greatly benefit from \pyMOR's
vectorized interface design. \mycode{NumpyVectorArray}, for instance, does so by calling
\numpy's \mycode{dot} method which is able to defer the task to highly optimized
BLAS implementations.
We expect similar performance benefits for external high-dimensional solvers,
when consecutive-in-memory arrays of vectors are available as native data
structures inside theses solvers.

\subsection{A large, nonlinear problem}\label{sec:burgers}

\begin{figure}[t]
  \footnotesize%
  \centering%
  \includegraphics[width=0.5\textwidth, trim=100 200 100 200, clip]{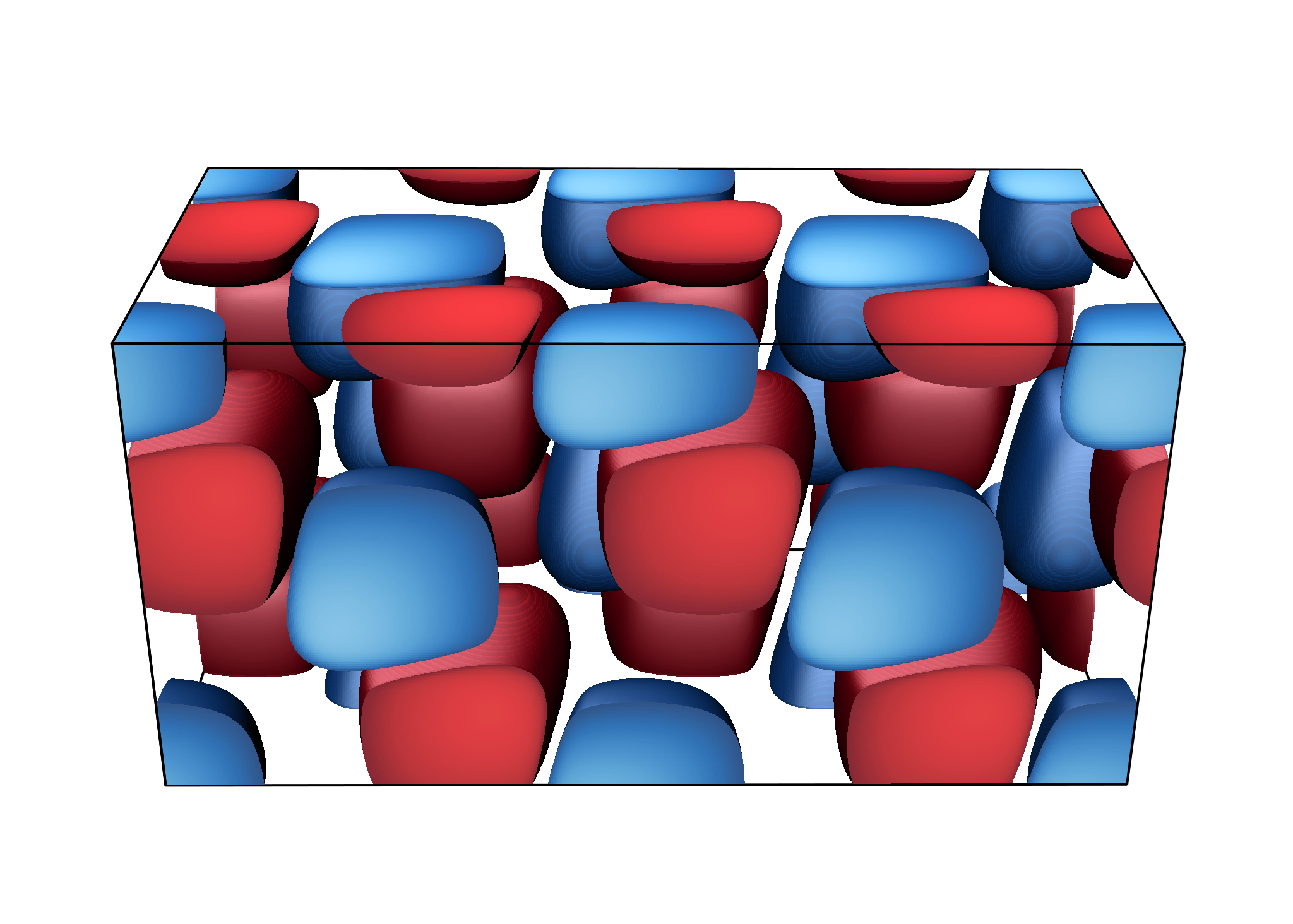}
  \hfill
  \includegraphics[clip, trim=0 674 448 0]{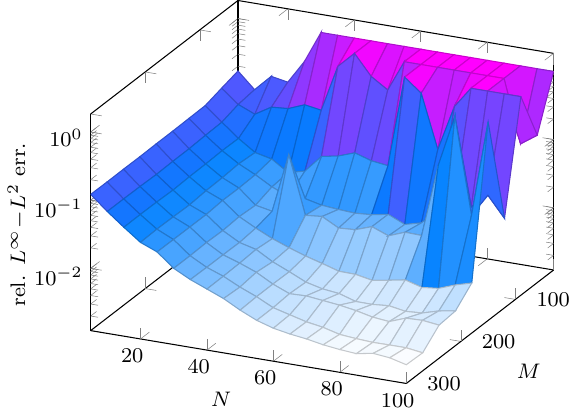}
  \caption{\emph{Left:} Plot of the solution of (\ref{eq:burgers}) for $\mu=2$
           at final time $t=0.3$. Only values in the intervals $[0, 0.4]$ (blue) and $[0.6, 1]$ (red) are displayed.
     \emph{Right:} Model reduction errors for RB approximation of
     (\ref{eq:burgers}). Maximum $L^\infty\!-L^2$-error on a test set of 10 random
     parameters for different reduced basis sizes $N$ and numbers of
     interpolation points $M$.
  }
  \label{fig:burgers_errors}
  \label{fig:burgers}
\end{figure}

To evaluate \pyMOR's ability to handle large-scale problems, we consider a
three-dimensional version of the Burgers-type problem already discussed in
\cite{DrohmannHaasdonkEtAl2012}, i.e., we solve the scalar conservation law
\begin{equation}\label{eq:burgers}
\begin{gathered}
	\partial_t u_\mu(t, x) + \nabla_x \cdot (\mathbf{v} \cdot u(t, x)^\mu) = 0,
	\qquad x \in [0, 2] \times [0, 1]^2, t \in [0, 0.3] \\
	u_\mu(0, x) = \frac{1}{2} ( 1 + \sin(2\pi x_1)\sin(2\pi x_2)\sin(2\pi
	x_3))
\end{gathered}
\end{equation}
for exponents $\mu \in [1, 2]$, periodic boundary conditions and constant
transport direction $\mathbf{v} = (1,  1, 1)$ (cf.\ Figure~\ref{fig:burgers}).
The problem was discretized using
a finite volume scheme on a $480 \times 240 \times 240$ voxel grid with 27.6
million degrees of freedom.

\begin{table}[t]
\footnotesize\centering
\caption{Time in seconds needed for the solution of (\ref{eq:burgers}) for a single
  parameter, using standalone \DUNE code in comparison to \DUNE code with time-stepping
  in \pyMOR (best of 3 runs).}
  \label{tab:burgers_times}
  \begin{tabular}{lrrrrrrrrr}
\toprule
\texttt{MPI} ranks & 1 & 2 & 3 & 6 & 12 & 24 & 48 & 96 & 192 \\
\midrule
\texttt{DUNE} &  16858 & 8532 & 5726 & 2959 & 1526 & 773 & 396 & 203 & 107 \\
\texttt{pyMOR} &  17683 & 8940 & 6050 & 3124 & 1604 & 815 & 417 & 213 & 110 \\
\midrule
overhead & 4.9\% & 4.8\% & 5.7\% & 5.6\% & 5.1\% & 5.4\% & 5.3\% & 4.9\% & 2.8\%\\
\bottomrule
  \end{tabular}
\end{table}

To keep the implementation, which is available in the supplementary material, as
simple as possible, we chose a basic Lax-Friedrichs numerical flux with
explicit Euler time discretization using 600 equidistant time steps. Our
\mycode{MPI}-parallel code only depends on the \DUNE grid interface and
includes hand-written \pyMOR bindings utilizing the \mycode{MPI} helper classes
and event loop covered in Section~\ref{sec:parallel}.

To evaluate the performance of the integration with \pyMOR, we compared the
solution time for (\ref{eq:burgers}) using a standalone version of the solver to
the time needed with time-stepping done by \pyMOR's
\mycode{explicit\textunderscore euler} time-stepper (Table~\ref{tab:burgers_times}). We observe that the \pyMOR integration shows very good
performance with only small overhead compared to the native \DUNE version.
All computations were performed on 1 to 16 nodes of the University of M\"unster's \textsc{PALMA} computing cluster.
Every node contains 48GB of main memory and two hexa-core Intel Xeon E5650 CPUs.

For the model order reduction we used the \textsc{EI-Greedy}~\cite{HaasdonkOhlbergerEtAl2008}
algorithm to generate the interpolation data for the nonlinear space
differential operator and a simple POD for the computation of the
reduced basis. In both cases, solution trajectories for 10 equidistant
parameters were chosen as input which had each been compressed beforehand using additional
PODs with a relative tolerance of $10^{-7}$. Thanks to the parallelization of
the high-dimensional discretization, the offline phase of the experiment could
be completed in only 3.4 hours.

Figure~\ref{fig:burgers_errors}
summarizes the approximation quality of the reduced model for different reduced
basis sizes and numbers of interpolation points. In particular we observe that
for a reduced basis of size 80 and 300 interpolation points, we can achieve a
relative $L^\infty\!-\!L^2$ model reduction error of $2.6\cdot10^{-3}$. A
solution for this reduced model takes on average $2.8$ seconds using a
single processor core, resulting in a speedup of 38 in comparison to a solution of
the high-dimensional model using all 192 cores or a speedup of 6000 in comparison
to a single-core computation.

\section*{Code availability} \pyMOR and all further code used for the production of the results in this work
are available under open source licenses.
The specific versions used here are included in the supplementary material.
Current versions of \pyMOR, including the wrapper classes for \mycode{FEniCS},
\mycode{deal.II} and \DUNE, as well as the code for the example in
Section~\ref{sec:burgers} can be found at
\url{http://www.pymor.org/}.

\section*{Acknowledgments}
This work has been supported by the German Federal Ministry of Education and Research (BMBF) under contract number 05M13PMA
and by the German Research Foundation (DFG) within the SPP 1648 \lq Software for Exascale Computing\rq\ program.


\begin{thebibliography}{10}

\bibitem{BallarinRozzaEtAl2015}
{\sc F.~Ballarin, G.~Rozza, and A.~Sartori}, {\em {RBniCS} - reduced order
  modelling in {FEniCS}}, ScienceOpen Posters,  (2015).

\bibitem{BangerthHartmannKanschat2007}
{\sc W.~Bangerth, R.~Hartmann, and G.~Kanschat}, {\em {deal.II} -- a general
  purpose object oriented finite element library}, ACM Trans. Math. Softw., 33
  (2007), pp.~24/1--24/27.

\bibitem{BBDEKO08B}
{\sc P.~Bastian, M.~Blatt, A.~Dedner, C.~Engwer, R.~Kl\"ofkorn, R.~Kornhuber,
  M.~Ohlberger, and O.~Sander}, {\em {A Generic Grid Interface for Parallel and
  Adaptive Scientific Computing. Part {II}: Implementation and Tests in
  {DUNE}}}, Computing, 82 (2008), pp.~121--138.

\bibitem{BBDEKO08A}
{\sc P.~Bastian, M.~Blatt, A.~Dedner, C.~Engwer, R.~Kl\"{o}fkorn, M.~Ohlberger,
  and O.~Sander}, {\em {A Generic Grid Interface for Parallel and Adaptive
  Scientific Computing. Part I: Abstract Framework}}, Computing, 82 (2008),
  pp.~103--119.

\bibitem{BelsonTuEtAl2014}
{\sc B.~A. Belson, J.~H. Tu, and C.~W. Rowley}, {\em Algorithm 945: Modred--a
  parallelized model reduction library}, ACM Trans. Math. Softw., 40 (2014),
  pp.~30:1--30:23.

\bibitem{BinevCohenEtAl2011}
{\sc P.~Binev, A.~Cohen, W.~Dahmen, R.~DeVore, G.~Petrova, and P.~Wojtaszczyk},
  {\em Convergence rates for greedy algorithms in reduced basis methods}, SIAM
  J. Math. Anal., 43 (2011), pp.~1457--1472.

\bibitem{BuhrEngwerEtAl2014}
{\sc A.~Buhr, C.~Engwer, M.~Ohlberger, and S.~Rave}, {\em A numerically stable
  a posteriori error estimator for reduced basis approximations of elliptic
  equations}, in Proceedings of the 11th World Congress on Computational
  Mechanics, X.~Oliver E.~Onate and A.~Huerta, eds., CIMNE, Barcelona, 2014,
  pp.~4094--4102.

\bibitem{BEOR15}
\leavevmode\vrule height 2pt depth -1.6pt width 23pt, {\em Arbilomod, a
  simulation technique designed for arbitrary local modifications}, arXiv
  e-prints,  (2015).
\newblock http://arxiv.org/abs/1512.07840.

\bibitem{ChaturantabutSorensen2010}
{\sc S.~Chaturantabut and D.~C. Sorensen}, {\em Nonlinear model reduction via
  discrete empirical interpolation}, SIAM J. Sci. Comput., 32 (2010),
  pp.~2737--2764.

\bibitem{DaversinVeysEtAl2013}
{\sc C.~Daversin, S.~Veys, C.~Trophime, and C.~Prud'homme}, {\em A reduced
  basis framework: Application to large scale non-linear multi-physics
  problems}, ESAIM: Proc., 43 (2013), pp.~225--254.

\bibitem{DeVorePetrovaEtAl2013}
{\sc R.~DeVore, G.~Petrova, and P.~Wojtaszczyk}, {\em Greedy algorithms for
  reduced bases in {B}anach spaces}, Constr. Approx., 37 (2013), pp.~455--466.

\bibitem{DrohmannHaasdonkEtAl2012a}
{\sc M.~Drohmann, B.~Haasdonk, S.~Kaulmann, and M.~Ohlberger}, {\em A software
  framework for reduced basis methods using dune-rb and rbmatlab}, in Advances
  in DUNE, Andreas Dedner, Bernd Flemisch, and Robert Klöfkorn, eds., Springer
  Berlin Heidelberg, 2012, pp.~77--88.

\bibitem{DrohmannHaasdonkEtAl2012}
{\sc M.~Drohmann, B.~Haasdonk, and M.~Ohlberger}, {\em Reduced basis
  approximation for nonlinear parametrized evolution equations based on
  empirical operator interpolation}, SIAM J. Sci. Comput., 34 (2012),
  pp.~A937--A969.

\bibitem{Haasdonk2013}
{\sc B.~Haasdonk}, {\em Convergence rates of the pod-greedy method}, ESAIM
  Math. Model. Numer. Anal., 47 (2013), pp.~859--873.

\bibitem{Ha14}
{\sc B.~Haasdonk}, {\em Reduced basis methods for parametrized {PDE}s -- {A}
  tutorial introduction for stationary and instationary problems}, 2016.
\newblock Chapter to appear in P. Benner, A. Cohen, M. Ohlberger and K.
  Willcox: "Model Reduction and Approximation: Theory and Algorithms", SIAM.

\bibitem{HaasdonkDihlmannEtAl2011}
{\sc B.~Haasdonk, M.~Dihlmann, and M.~Ohlberger}, {\em A training set and
  multiple bases generation approach for parameterized model reduction based on
  adaptive grids in parameter space}, Math. Comput. Model. Dyn. Syst., 17
  (2011), pp.~423--442.

\bibitem{HaasdonkOhlbergerEtAl2008}
{\sc B.~Haasdonk, M.~Ohlberger, and G.~Rozza}, {\em A reduced basis method for
  evolution schemes with parameter-dependent explicit operators}, Electron.
  Trans. Numer. Anal., 32 (2008), pp.~145--161.

\bibitem{HesthavenRozzaEtAl2016}
{\sc J.S. Hesthaven, G.~Rozza, and B.~Stamm}, {\em Certified Reduced Basis
  Methods for Parametrized Partial Differential Equations}, SpringerBriefs in
  Mathematics, Springer International Publishing, 2016.

\bibitem{JohnsenTaylorEtAl2015}
{\sc S.F. Johnsen et~al.}, {\em Niftysim: A gpu-based nonlinear finite element
  package for simulation of soft tissue biomechanics.}, Int J Comput Assist
  Radiol Surg, 10 (2015), pp.~1077--1095.

\bibitem{JonesOliphantEtAl2001}
{\sc E.~Jones, T.~Oliphant, P.~Peterson, et~al.}, {\em {SciPy}: Open source
  scientific tools for {Python} (\mycode{http://www.scipy.org/})}, 2001--2015.

\bibitem{KnezevicPeterson2011}
{\sc D.~J. Knezevic and J.~W. Peterson}, {\em A high-performance parallel
  implementation of the certified reduced basis method}, Computer Methods in
  Applied Mechanics and Engineering, 200 (2011), pp.~1455 -- 1466.

\bibitem{LoggMardalEtAl2012}
{\sc A.~Logg, K.-A. Mardal, G.N. Wells, et~al.}, {\em Automated Solution of
  Differential Equations by the Finite Element Method}, Springer, 2012.

\bibitem{LoggWells2010}
{\sc A.~Logg and G.N. Wells}, {\em Dolfin: Automated finite element computing},
  ACM Trans. Math. Softw., 37 (2010), pp.~20:1--20:28.

\bibitem{ORS16}
{\sc M.~Ohlberger, S.~Rave, and F.~Schindler}, {\em Model reduction for
  multiscale lithium-ion battery simulation}, in ENUMATH 2015, Ankara, Turkey,
  LNCSE, LNCSE, Springer, 2016.

\bibitem{OhlbergerRaveEtAl2014}
{\sc M.~Ohlberger, S.~Rave, S.~Schmidt, and S.~Zhang}, {\em A model reduction
  framework for efficient simulation of li-ion batteries}, in Finite Volumes
  for Complex Applications VII-Elliptic, Parabolic and Hyperbolic Problems,
  J\"urgen Fuhrmann, Mario Ohlberger, and Christian Rohde, eds., Springer,
  2014, pp.~695--702.

\bibitem{OS2015}
{\sc M.~Ohlberger and F.~Schindler}, {\em Error {C}ontrol for the {L}ocalized
  {R}educed {B}asis {M}ultiscale {M}ethod with {A}daptive {O}n-{L}ine
  {E}nrichment}, SIAM J. Sci. Comput., 37 (2015), pp.~A2865--A2895.

\bibitem{Oliphant2007}
{\sc T.~E. Oliphant}, {\em Python for scientific computing}, Computing in
  Science \& Engineering, 9 (2007), pp.~10--20.

\bibitem{PateraRozza}
{\sc A.~T. Patera and G.~Rozza}, {\em Reduced basis approximation and a
  posteriori error estimation for parametrized partial differential equations,
  version 1.0,}.
\newblock Copyright MIT 2006--2007, to appear in (tentative rubric) MIT
  Pappalardo Graduate Monographs in Mechanical Engineering.

\bibitem{PerezGranger2007}
{\sc F.~P\'erez and B.~E. Granger}, {\em {IP}ython: a system for interactive
  scientific computing}, Computing in Science and Engineering, 9 (2007),
  pp.~21--29.

\bibitem{QuarteroniManzoniEtAl2016}
{\sc A.~Quarteroni, A.~Manzoni, and F.~Negri}, {\em Reduced Basis Methods for
  Partial Differential Equations}, La Matematica per il 3+2, Springer
  International Publishing, 2016.

\bibitem{dunegdt}
{\sc F.~Schindler and R.~Milk}, {\em \dune{gdt}
  (\mycode{http://dx.doi.org/10.5281/zenodo.45465})}, 2015.

\end{thebibliography}
\end{document}